\documentclass[acmtog]{acmart}
\hypersetup{draft}
\newcommand{\revised}[1]{{\color{black} #1}}
\usepackage[ruled]{algorithm2e} 

\SetAlFnt{\small}
\SetAlCapFnt{\small}
\SetAlCapNameFnt{\small}
\SetAlCapHSkip{0pt}

\definecolor{mycolor_dashline}{RGB}{213,94,0}%
\usepackage{amsmath,bm}
\usepackage{graphicx}
\usepackage{color}
\usepackage{multirow}
\usepackage{overpic}
\def\etal{\emph{et al.}}
\usepackage{enumitem}

\usepackage{pgfplots}
\usepackage{pgf,tikz}
\usetikzlibrary{matrix,backgrounds}

\setcopyright{acmcopyright}
\acmJournal{TOG}
\acmYear{2020}\acmVolume{39}\acmNumber{4}\acmArticle{122}\acmMonth{7} \acmDOI{10.1145/3386569.3392443}

\begin{document}
	
	\title{MGCN: Descriptor Learning using Multiscale GCNs}

	\author{Yiqun Wang}
	\affiliation{%
		\institution{NLPR, Institute of Automation, CAS; School of AI, University of Chinese Academy of Sciences; KAUST}
	}
	 \email{yiqun.wang@nlpr.ia.ac.cn}
	\author{Jing Ren}
	\affiliation{
		\institution{KAUST}
	}
	 \email{jing.ren@kaust.edu.sa}
	\author{Dong-Ming Yan}
	\authornote{Corresponding author.}
	\affiliation{%
		\institution{NLPR, Institute of Automation, CAS; School of AI, University of Chinese Academy of Sciences}
	}
	 \email{yandongming@gmail.com}
	\author{Jianwei Guo}
	\affiliation{%
		\institution{NLPR, Institute of Automation, CAS; School of AI, University of Chinese Academy of Sciences}
	}
	 \email{jianwei.guo@nlpr.ia.ac.cn}
	\author{Xiaopeng Zhang}
	\affiliation{%
		\institution{NLPR, Institute of Automation, CAS; School of AI, University of Chinese Academy of Sciences}
	}
	 \email{xiaopeng.zhang@ia.ac.cn}
	\author{Peter Wonka}
	\affiliation{
		\institution{KAUST}
	}
	 \email{pwonka@gmail.com}

	\begin{abstract}
		We propose a novel framework for computing descriptors for characterizing points on three-dimensional surfaces.
		First, we present a new non-learned feature that uses graph wavelets to decompose the Dirichlet energy on a surface. We call this new feature \emph{Wavelet Energy Decomposition Signature} (WEDS).
		Second, we propose a new \emph{Multiscale Graph Convolutional Network} (MGCN) to transform a non-learned feature to a more discriminative descriptor.
		Our results show that the new descriptor WEDS is more
		discriminative than the current state-of-the-art non-learned descriptors and that the combination of WEDS and MGCN is better than the state-of-the-art learned descriptors.
		An important design criterion for our descriptor is the robustness
		to different surface discretizations including triangulations with varying numbers of vertices.
		Our results demonstrate that previous graph convolutional networks significantly overfit to a particular resolution or even a particular triangulation, but MGCN generalizes well to different surface discretizations.
		In addition, MGCN is compatible with previous descriptors and it can also be used to improve the performance of other descriptors, such as the heat kernel signature, the wave kernel signature, or the local point signature.
	\end{abstract}
	
	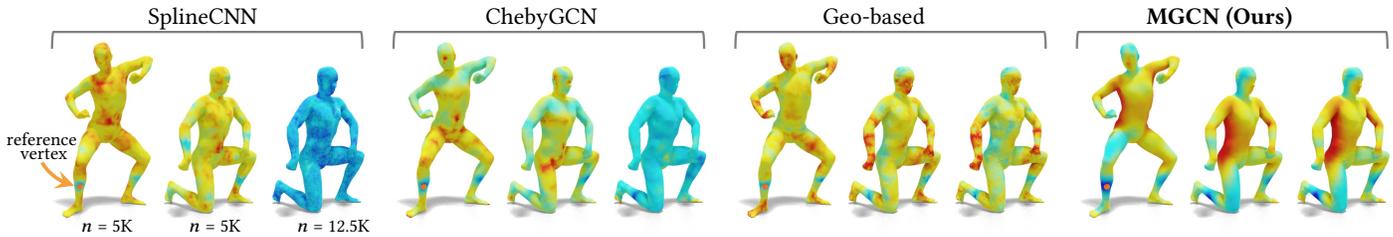
\begin{teaserfigure}
		\centering
		\begin{overpic}[trim=0cm -1cm 0cm -2cm,clip,width=1\linewidth,grid=false]{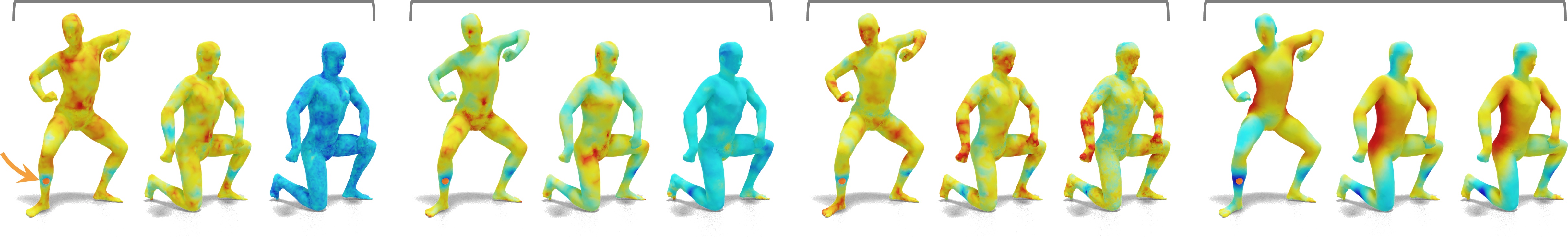}
			\put(-2.5,7.4){\footnotesize reference}
			\put(-1.5,6.4){\footnotesize vertex}
			\put(8,16.4){SplineCNN}
			\put(33,16.4){ChebyGCN}
			\put(58,16.4){Geo-based}
			\put(82,16.4){\textbf{MGCN (Ours)}}
			\put(3,1){\footnotesize $n$ = 5K}
			\put(11,1){\footnotesize $n$ = 5K}
			\put(19,1){\footnotesize $n$ = 12.5K}
		\end{overpic}\vspace{-12pt}
		\caption{Visualization of dissimilarity maps. For a vertex highlighted by the orange arrow, we take \revised{its} learned desriptor and visualize the difference to other vertex descriptors on the same shape, another shape with the same 5K resolution, and the other shape in a different 12.5K resolution. We compare four different learned descriptors, from left to right: SplineCNN, ChebyGCN, Geodesic-based method~\cite{Wang_2019_CVPR}, and MGCN. All networks are trained on 5K and tested on 5K and 12.5K resolution. We can see that our network MGCN is most consistent between different resolutions.}
		\label{fig:overfits}
	\end{teaserfigure}

	%
	%


	\maketitle

	\section{Introduction}
	
	Designing descriptors for surface points is a fundamental problem in geometry processing as descriptors are a building block for many applications, such as shape matching, registration, segmentation, and retrieval.
	
	A good descriptor should satisfy two criteria: (1) The descriptor should be \emph{discriminative} to map similar surface points to similar values and dissimilar surface points to dissimilar values. The definition of similarity depends on the application. In our setting, we consider the very popular requirement that descriptors should be invariant to rigid and near-isometric deformations of the surface.
	(2) The descriptor should be \emph{robust} to different discretizations of the surface, \emph{e.g}., meshes of different resolution and triangulation. If the descriptor discriminates surface points based on the discretization, we also say it overfits or lacks generalization.
	
	Generally, we can distiguish two types of descriptor computation: supervised and \revised{non-learned}. Examples of \revised{non-learned} descriptors are the \emph{Wave Kernel Signature} (WKS) and the \emph{Heat Kernal Signature} (HKS). While these descriptors are robust to different surface discretization, there is a lot of room for improvement making them more discriminative. This can been done successfully using neural networks to compute supervised descriptors. A very promising type of network architecture are graph convolutional networks, such as chebyGCN~\cite{defferrard2016convolutional}, GCN~\cite{kipf2017semi}, SplineCNN~\cite{fey2018splinecnn}, and DGCNN~\cite{wang2019dynamic}. Even though many of these networks have not been applied to descriptor learning directly, their adaption to descriptor learning only requires little effort. However, the current state of the art is typically not robust to different surface discretizations and overfits. See Fig.~\ref{fig:overfits} for an illustration. One main reason for this overfitting is that the convolution operation typically depends on a $k$-ring neighborhood of a surface point. This changes the spatial support of a convolutional filter if, for example, the resolution of the underlying surface triangulation changes. One possible approach to make descriptor learning robust to different resolutions is to resample the surface. This approach has other drawbacks, such as the additional complexity of resampling and the loss of information reducing the discrimination performance.

	In this paper, we propose contributions to \revised{non-learned} and supervised descriptor computation leveraging the power of wavelets. For the learning part, we introduce a novel graph convolutional network called \emph{Multiscale Graph Convolutional Network} (MGCN). Our results will show significant improvements to descriptor performance even when tested under a variety of different surface discretizations. The key novelty of our network is a convolution operation expressed in the wavelet basis. This lets us define a multiscale convolution with filters of both local and global support.
	
	For the \revised{non-learned} descriptor computation part, we theoretically derive a novel local spectral feature called \emph{Wavelet Energy Decomposition Signature} (WEDS) from the Dirichlet energy. Different from traditional spectral descriptors (\emph{e.g}., \emph{Global Point Signature} (GPS), HKS, and WKS), we introduce additional vertex coordinate information to capture more distinctive attributes. Compared with \emph{Local Point Signature} (LPS)~\cite{Wang_2019_CVPR}, our new descriptor uses wavelets to capture both local and global information, which is more discrimintive.
	
	Our extensive experimental evaluations indicate that the WEDS descriptor outperforms recent state-of-the-art \revised{non-learned} descriptors. Further, WEDS can be combined with MGCN to improve upon the currently best supervised descriptors. Besides the traditional evaluation of descriptor performance with respect to rigid, near-isometric, and non-isometric surface deformations, we also evaluate the robustness to different surface discretizations.

	The main contributions of this work are as follows:
	\begin{itemize}
		\item We design a new graph convolutional network named \emph{Multiscale Graph Convolutional Network}. While we focus on descriptor learning as main application, the robustness to resolution of our new convolution layer holds promise for many additional applications.
		\item We present a novel multiscale feature called \emph{Wavelet Energy Decomposition Signature} based on energy decomposition that improves upon state-of-the-art \revised{non-learned} descriptors.
	\end{itemize}
	
	\section{Related Work}  
	\label{sec:relatedWork}
	
	We review related work for descriptor computation in three categories and subsequently related work for optimization-based matching methods and graph convolutional neural networks.
	
	\subsection{Descriptor Generation}
	\paragraph{Spatial domain approaches.}
	Descriptors directly constructed in the spatial domain often rely on histograms. \emph{Spin Images} (SI)~\cite{johnson1999using} and \emph{3D Shape Context} (3DSC)~\cite{frome2004recognizing} are generated by creating accumulators, which divide the local space into different bins and calculate the number of points that fall into each bin. \emph{Signature of Histogram of Orientations} (SHOT)~\cite{tombari2010unique} is constructed by accumulating the normal angles of the key and neighboring points in the neighborhood space. Unlike the SHOT descriptor, the \emph{Mesh Histogram of Oriented Gradients} (MeshHOG)~\cite{zaharescu2009surface} descriptor is another histogram based on the orientations of the gradients on the mesh. The \emph{Rotational Projection Statistics} (RoPS)~\cite{guo2013rotational} descriptor is generated by rotationally projecting neighboring points onto 2D planes and calculating a set of statistics. Spatial domain descriptors generally have the following performance characteristics. First, they heavily rely on local information, but do not capture global information. Second, the descriptors are sensitive to the discretization of the surface. While this is desirable for some applications, an important goal of our work is to be robust to different surface discretizations.
	
	\paragraph{Spectral domain approaches.}
	Many spectral descriptors have been proposed to deal with isometric deformations. Especially popular are intrinsic descriptors based on the Laplace-Beltrami operator.
	Shape-DNA~\cite{reuter2006laplace} considers the spectrum of the Laplace-Beltrami operator as the descriptor because the spectrum is isometry-invariant and independent of spatial position.
	GPS~\cite{rustamov2007laplace} combines the spectrum and eigenfunctions to obtain a descriptor on each vertex.
	HKS~\cite{sun2009concise}, scale-invariant HKS~\cite{bronstein2010scale}, and WKS~\cite{aubry2011wave}  are proposed based on diffusion geometry.
	The intrinsic properties make the descriptors invariant to isometric deformation. LPS~\cite{Wang_2019_CVPR} combines coordinate information and intrinsic geometric information to get a more robust descriptor. The \emph{Discrete Time Evolution Process} (DTEP) descriptor~\cite{melzi2018discrete} focuses on non-isometric deformations and achieved better results. But these methods take a lot of time to compute geodesic distances or solve optimization problems. In our results, we compare to the best performing descriptors to demonstrate an important improvement in performance. Our discussion in Section~\ref{subsec:discussionofWEDS} will explain why we are able to beat the state of the arts in more detail.
	
	\paragraph{Deep learning approaches.}
	We call the descriptors reviewed in the previous two paragraphs \revised{non-learned}. By contrast, supervised descriptors use supervised learning, mainly deep learning, to extract shape descriptors. Wei~\etal~\shortcite{wei2016dense} generate descriptors by using  a large dataset of depth maps for training.
	Huang \etal~\shortcite{huang2018learning} extract local descriptors by training on multiple rendered views in multiple scales.
	Zeng \etal~\shortcite{zeng20173dmatch} use 3D volumetric convolutional neural networks to generate local descriptors for robustly matching RGB-D data.
	The method of \emph{Compact Geometric Features} (CGF)~\cite{khoury2017learning} maps high-dimensional histograms into a low-dimensional Euclidean space to generate descriptors on unstructured point clouds. Deng \etal~\shortcite{deng2018ppfnet} find matches in unorganized point clouds by adapting the PointNet architecture.
	Although these methods have obtained good results through learning, they do not make full use of the structure of 3D data. Multi-view-based methods must solve the problem of occlusion. Voxel-based methods consume a lot of resources and cannot explore the details of an object. Learning methods that work by feeding point cloud coordinates or histogram features into multi-layer perceptrons are currently not able to extract enough important information from the data.
	
	Some other descriptors are generated by exploiting the correlation between local points in the frequency domain or spatial domain.
	\emph{Optimal Spectral Descriptors} (OSD)~\cite{litman2014learning} are constructed by learning parametric filters in the spectral domain. Boscaini~\etal~\shortcite{boscaini2015learning} generalize the windowed Fourier transform to learn local shape descriptors on manifolds. Anisotropic diffusion descriptors~\cite{boscaini2016anisotropic} based on anisotropic diffusion are constructed by using a fully connected neural network to learn the kernel filters.
	In the spatial domain, Masci \etal~\shortcite{masci2015geodesic} design a geodesic convolutional network to learn shape descriptors on manifolds by extracting and regularly charting geodesic local patches and designing a patch operator for convolution.
	Wang \etal~\shortcite{wang2018learning} and Guo \etal~\shortcite{guo2020learning} employ a deep learning framework by projecting geodesic local patches into local geometry images to learn descriptors, respectively. Subsequently, LPS~\cite{Wang_2019_CVPR} is proposed on geodesic local patches and used with deep learning to construct a more discriminative descriptor. Although these spatial domain methods can convolve local regions on manifolds and get better results, they need to extract local geodesic patches, which is very time-consuming. In addition, it is very difficult to maintain a disk topology if the local region becomes larger and the surface contains topological holes.
	
	\subsection{Optimization-based Shape Matching Methods}
	\revised{Optimization-based matching approaches optimize a dense map between a pair of shapes. Some of these methods are initialized by shape descriptors. We focus on the related state-of-the-art methods and refer the reader to a recent survey~\cite{biasotti2016recent} for more details on shape matching. For near-isometric shapes, one of the most successful approaches is based on functional maps~\cite{ovsjanikov2012functional}. For each shape, a Laplace-Beltrami basis is computed to express the function space. The functional map between two shapes is encoded as a matrix. Deep functional maps~\cite{litany2017deep} combine functional maps with deep learning. In the proposed network, the initial stage learns a shape descriptor that is trained with the goal of being useful in shape matching. Halimi~\etal~\shortcite{halimi2019unsupervised} extend deep functional maps so that they can be trained in an unsupervised manner. For non-isometric shapes, \emph{Blended Intrinsic Map} (BIM)~\cite{kim2011blended} is the current state of the art where a set of maps are proposed and then blended together to output a single map. }

	\subsection{Graph Convolutional Networks}
	Recently, a large number of graph convolutional learning methods have emerged. This learning method has achieved high performance on irregular data. Spectral CNN~\cite{Bruna2014Spectral} is the first to perform convolution operations on the graph through a frequency domain transform. ChebyGCN~\cite{defferrard2016convolutional} simplifies spectral CNN by designing spectral filters using a $k$-order polynomial parametrization. GCN~\cite{kipf2017semi} further simplifies polynomials to $1$-order and is suitable for semi-supervised learning. SplineCNN~\cite{fey2018splinecnn} uses B-Spline kernels to weight the relationship between a point and its neighborhood. Although this network has strong fitting ability, the generalization is not strong as the pseudo-coordinates on the edges are not invariant to rigid transformations. \revised{\emph{Feature-Steered Network} (FeaStNet)~\cite{verma2018feastnet} proposes a learnable matrix to weight pseudo-coordinates in $k$-ring neighbors over the graph.} Wang \etal~\shortcite{wang2019dynamic} present a \emph{Dynamic Graph CNN} (DGCNN) on point cloud, which can build dynamic connections by selecting a $k$-neighborhood in feature space. \revised{ SpiralNet~\cite{lim2018_correspondence_learning} and SpiralNet++~\cite{gong2019spiralnet++} define spiral convolution for a $k$-ring neighborhood on the mesh. The convolution operator of MeshCNN~\cite{hanocka2019meshcnn} is defined on the four edges of the two incident triangles of an
		edge. } The problem with the graph convolution using \revised{ a $1$-neighborhood, a $k$-neighborhood or a $k$-ring neighborhood} is that it is not suitable for inputs of different resolutions because the size of the receptive field changes depending on the discretization of the input. The \emph{Graph Wavelet Neural Network} (GWNN)~\cite{XuSCQC19} uses the wavelet transform to formulate convolutions on a graph. The problem with this convolution method is that only a single-scale wavelet is used in the transformation. In addition, the number of filters in this algorithm is related to the number of vertices, so that convolution in multiple resolutions cannot be achieved.
	
	Even though there are many methods of graph convolutional networks, a graph convolutional network is rarely used to learn shape descriptors. One of the problems is that graph convolutions are strongly influenced by the neighborhood relationships stemming from the surface discretization. Therefore, the result is overly sensitive to the discretization of the surface.
	We focus on the issues of resolution and triangulation in graph convolutional networks and present a novel network to generate an informative descriptor that is robust to the change of resolution and triangulation.

	\section{Problem Statement and Overview}  
	
	Given is a mesh $\mathcal{M}$ as discretization of an underlying smooth two-manifold surface defined as $(V, E)$, where $V=\{v_i|i=1,...,N\}$ and $E$ are the sets of vertices and edges, respectively. The vertex coordinates are defined by the function $\pmb{\rm{X}}=(\pmb{\rm{x}}_1, \pmb{\rm{x}}_2, \pmb{\rm{x}}_3): V\to \mathbb{R}^3$
	Our goal is to compute a local descriptor ${\rm{f}}(v_i) \in \mathbb{R}^d$ for any given vertex $v_i$.
	
	The local descriptor is generated in two stages: \revised{non-learned} descriptor computation and supervised descriptor learning.
	At the first stage, we compute our proposed descriptor WEDS for each vertex in the wavelet domain that is robust to the change of resolution, triangulation, scale, and rotation. The second stage is about descriptor learning. Inspired by the derivation of WEDS, we propose a graph convolutional network called MGCN to generate better descriptors from WEDS. Benefiting from the expressiveness of graph wavelets, our network can be trained on one resolution and tested on other resolutions without significant reduction of performance.
	
	\section{Non-learned Descriptor Computation}  
	
	In this section, we first review Laplacian eigenfunctions and graph wavelets. Then, we propose a new type of shape descriptor, WEDS. Finally, we discuss properties and advantanges of WEDS.
	
	\subsection{Laplacian Eigenfunctions}      
	
	Let $\mathcal{S}$ denote a continuous surface. We can find an orthonormal basis on $\mathcal{S}$, containing the $k$ smoothest possible functions that are orthogonal to each other, by finding the first $k$ eigenfunctions of $\Delta$~\cite{bronstein2017geometric}, the Laplace-Beltrami operator:
	\begin{equation}
	\Delta {\mathbf{\phi} _i} = {\lambda _i}{\mathbf{\phi} _i},i = 0,1,...,k-1,
	\label{con:LB_eigen}
	\end{equation}
	where $\left\{ {{\lambda _i}|i = 0,1,...,k-1} \right\}$ are the smallest $k$ eigenvalues in increasing order.
	To simplify the notation, we use the same variable names for discrete and continuous settings. The difference should be clear from the context.

	In the discrete setting of a triangulated mesh $\mathcal{M}$ with $N$ vertices, we can discretize the Laplace-Beltrami operator as follows:
	\begin{equation}
	\mathbf{L}\bm{{\phi}_i} = {\lambda _i}\mathbf{A}\bm{{\phi }_i},i = 0,1,...,k - 1,
	\label{con:LBM_eigen}
	\end{equation}
	where $\mathbf{L}$ is a standard cotangent Laplacian matrix with size $N\times N$, $\mathbf{A}$ is the $N\times N$ diagonal area matrix. $\bm{\phi}_i$ is a $N\times 1$ vector, the eigenvector with respect to the eigenvalue $\lambda_i$.
	Also, note that for the generalized eigenvalue problem, the eigenvectors $\bm{\phi}_i$ are orthogonal to each other in terms of the $\mathbf{A}$-dot product: ${\left\langle {\bm{{\phi}_i},\bm{{\phi}_j}} \right\rangle _\mathbf{A}} = \bm{\phi}_i^T\mathbf{A}\bm{{\phi}_j}$.
	Any function $f$ defined on a smooth surface can be expressed as a weighted combination of the eigenfunctions: $f = \sum\nolimits_{j=0}^{\infty}{{\sigma_j}{\mathbf{\phi}_j}}$,
	where $\sigma_j$ is the coefficient corresponding to the $j^{\text{th}}$ eigenfunction.
	In the discrete case, the coefficients can be calculated by
	\begin{equation}
	{\sigma _j} = {\left\langle {\pmb{\rm{f}},\bm{{\phi} }_j} \right\rangle _\mathbf{A}} = {\pmb{\rm{f}}^T}\mathbf{A}\bm{{\phi }_j},
	\label{con:coeff}
	\end{equation}
	where $\pmb{\rm{f}}$ is the corresponding discretization of a smooth function $f$ that is defined on the vertices of a triangulated mesh.
	
	We can see that with the help of basis functions, a function on the surface can be transformed into a set of coefficients. Graph wavelets, explained next, make use of the same idea. The main difference is that instead of eigenfunctions of the Laplace-Beltrami operator, wavelet and scaling functions are used as basis functions.

	\subsection{Graph Wavelets}        
	
	We build on the graph wavelet framework described in paper by Hammond et al.~\shortcite{hammond2011wavelets}. To make the application specific to meshes using the cotangent Laplacian  rather than the uniform Laplacian we build on the notation of  ~\cite{masoumi2017spectral} where the inner product is defined with respect to the area matrix $\mathbf{A}$.
	
\begin{figure}[!t]
		\centering
		\begin{overpic}[trim=0cm 0cm 0cm -3cm,clip,width=0.99\linewidth,grid=false]{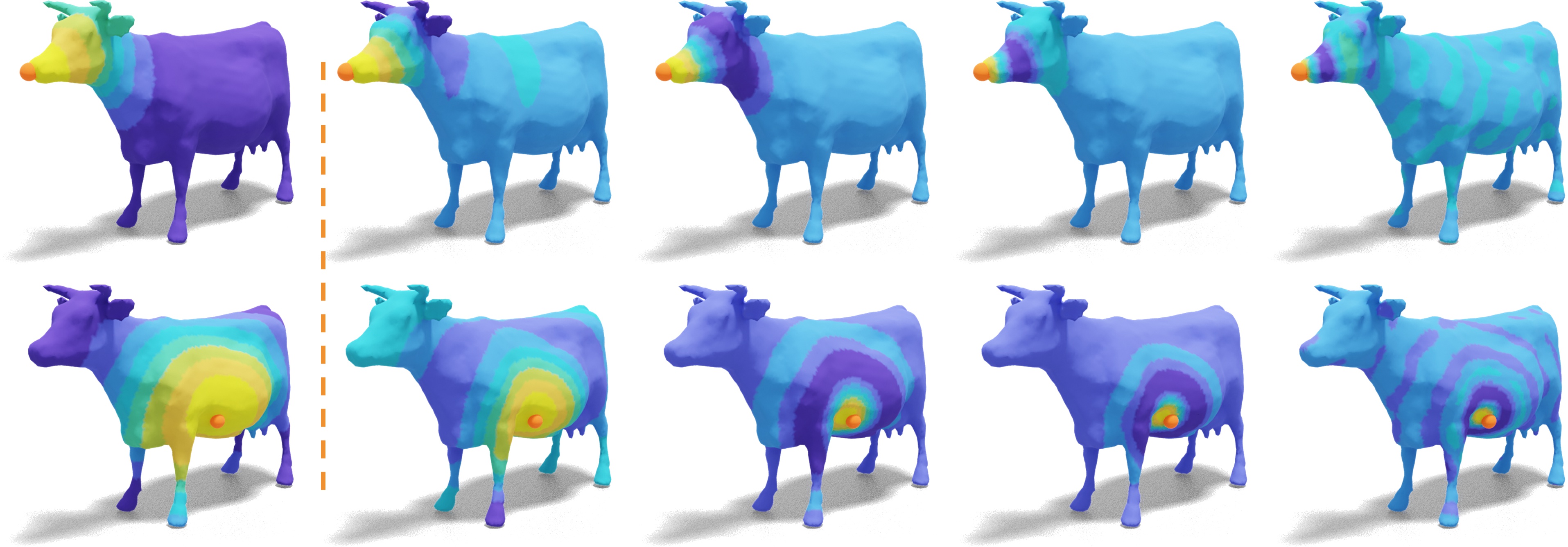}
			\put(0,37){\footnotesize Scaling function}
			\put(35,37){\footnotesize Wavelet functions with different time scales}
		\end{overpic}\vspace{-10pt}
		\caption{Illustration of wavelet functions and scaling functions. The top row corresponds to a vertex on the mouth and the bottom row to a vertex on the stomach (shown as orange spheres). The first column shows the scaling functions $\bm{\xi} _{v}$ corresponding to the vertices. The second to fifth columns correspond to wavelet functions $\bm{\psi}_{t,v}$ of different time scales t = 0.0210, t= 0.0102, t= 0.0024, and t=5.59$e^{-4}$.}
		\label{fig:wavelet_example_vis}\vspace{-8pt}
	\end{figure}

	\paragraph{Wavelet function.} One graph wavelet function $\bm{{\psi} _{t,v}}$ is defined per vertex $v$ per time scale $t$. We denote the number of time scales as $K$ ($K$ is typically less than 100).
	To construct a wavelet function, a filter function $g$ is used. Examples for $g$ are the Mexican hat or cubic splines. Let $a\left( v \right)$ be the Voronoi area at vertex $v$ and $\phi _j\left( v \right)$ be the element of vector $\bm{\phi}_j$ corresponding to vertex $v$.  Then $\bm{{\psi} _{t,v}}$, the spectral graph wavelet localized at vertex $v$ and scale $t$, is given by
	\begin{equation} 
	\bm{{\psi} _{t,v}} = \sum\limits_{j{\rm{ = 0}}}^{N - 1} {a\left( v \right) g\left( {t{\lambda _j}} \right){\phi _j}\left( v \right)\bm{{\phi} _j}}. 
	\label{con:WB}
	\end{equation}
	We show examples in Fig.~\ref{fig:wavelet_example_vis} to demonstrate the local property of wavelets.
	
	To project a given function $\pmb{\rm{f}}$ onto the wavelet basis, we compute the inner product between the wavelet functions and the given function $\pmb{\rm{f}}$.
	The spectral graph wavelet coefficients are then defined as:
	\begin{equation}
	{W_{\pmb{\rm{f}}}}\left( {t,v} \right) = {\left\langle {{\pmb{\rm{f}},\bm{\psi} _{t,v}}} \right\rangle _{\mathbf{A}}} = \sum\limits_{j{\rm{ = 0}}}^{N - 1} {a\left( v \right) g\left( {t{\lambda _j}} \right){\sigma _j}{\phi _j}\left( v \right)}.
	\label{con:WFT}
	\end{equation}
	
	\paragraph{Scaling function.} In addition to the graph wavelet functions, there is a single scaling function $\bm{{\xi} _{v}}$ defined per vertex $v$ on the surface. The scaling function is defined via a filter function $h(x)$. Typically, $h(x)$ is the same type of function as $g(x)$, but using different parameters. The scaling function captures low-frequency information and is given by
	\begin{equation} 
	\bm{{\xi} _{v}} = \sum\limits_{j{\rm{ = 0}}}^{N - 1} {a\left( v \right) h\left( {{\lambda _j}} \right){\phi _j}\left( v \right)\bm{{\phi} _j}}. 
	\label{con:SWB}
	\end{equation}
	An illustration is shown in Fig.~\ref{fig:wavelet_example_vis}.
	Similar to the wavelet functions, we compute the inner product between a given function $\pmb{\rm{f}}$ and the scaling functions to obtain the scaling function coefficients
	\begin{equation}
	{S_{\pmb{\rm{f}}}}\left( {v} \right) = \left\langle {{\pmb{\rm{f}},\bm{\xi} _{v}}} \right\rangle _{\mathbf{A}} = \sum\limits_{j{\rm{ = 0}}}^{N - 1} {a\left( v \right) h\left( {{\lambda _j}} \right){\sigma _j}{\phi _j}\left( v \right)}.
	\label{con:SWFT}
	\end{equation}
	
	To use graph wavelets in our framework, the filter functions $g(x)$ and $h(x)$ cannot be chosen arbitrarily, as we require that they form a Parseval frame~\cite{stankovic2019vertex}. This is necessary so that the coefficients can be used to recover the original signal in the discrete case as follows:
	\begin{equation} 
	\small
	\pmb{\rm{f}} = \sum\limits_{m{\rm{ = 1}}}^{K} {\sum\limits_v {a\left( v \right)^{-1}{W_{\pmb{\rm{f}}}}\left( {t_m,v} \right)\bm{{\psi} _{t_m,v}}} }  + \sum\limits_v {a\left( v \right)^{-1}{S_{\pmb{\rm{f}}}}\left( v \right)\bm{{\xi} _v}} ,  
	\label{con:CON}
	\end{equation}
	where $t_m$ is the $m^{th}$ scale of the wavelet function.
	We will discuss Parseval frames and the choice of filter functions in Section~\ref{sec:Multiscale_Filters}.
	
	The above formula can be simplified as
	\begin{equation} 
	\pmb{\rm{f}} = \sum\limits_{m{\rm{ = 0}}}^{K} {\sum\limits_v {a\left( v \right)^{-1}{W_{\pmb{\rm{f}}}}\left( {t_m,v} \right)\bm{{\psi} _{t_m,v}}} },  
	\label{con:CONS}
	\end{equation}
	where ${W_{\pmb{\rm{f}}}}\left( {t_0,v} \right) = {S_{\pmb{\rm{f}}}}\left( v \right)$, and $\bm{{\psi} _{t_0,v}} = \bm{{\xi} _v}. $ The proof is given in the appendix.
	
	
	\begin{figure}[!t]
		\centering
		\begin{overpic}[trim=1cm 0cm 0cm -3cm,clip,width=1\linewidth,grid=false]{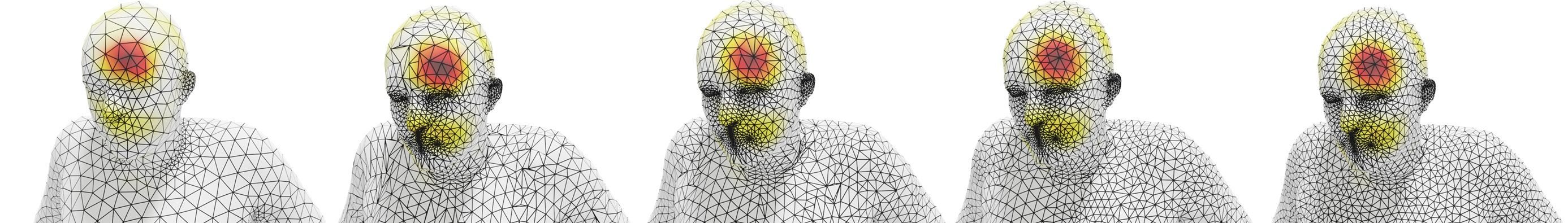}
			\put(4,16){\footnotesize $n=5K$}
			\put(23,16){\footnotesize $n=8K$}
			\put(42,16){\footnotesize $n=10K$}
			\put(62,16){\footnotesize $n=12K$}
			\put(83,16){\footnotesize $n=15K$}
		\end{overpic}\vspace{-6pt}
		\caption{We choose five resolutions to show the wavelet function on one vertex. From left to right: 5K, 8K with random filpping, 10K, 12K, 15K.  The model with 5K vertices is remeshed and 8K is with random filpping. From the illustration, the wavelet functions are robust with respect to change of resolution and triangulation.}
		\label{fig:wavelet_local}\vspace{-8pt}
	\end{figure}
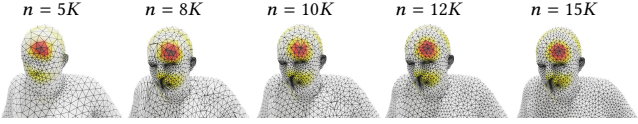
	
	\subsection{Wavelet Energy Decomposition Signature}        
	
	To derive our new descriptor, we combine multiple ideas. First, we would like to start from the coordinate functions $\pmb{\rm{X}}=(\pmb{\rm{x}}_1, \pmb{\rm{x}}_2, \pmb{\rm{x}}_3)$, because they completely describe the shape and are therefore very informative. Second, to make this information invariant to rigid transformations, we employ the sum of the Dirichlet energy of the three coordinate functions. This summation equals to the surface area and the computation of the Dirichlet energy is generally robust to different discretizations. Third, to aggregate local information in an area around the vertex, we employ graph wavelets at different scales described previously. This ensures that our descriptor is more discriminative than current state-of-the-art descriptors.
	
	Given a smooth real-valued function $f: \mathcal{S}\to \mathbb{R}$ defined on the surface, the Dirichlet energy measures how smooth the function $f$ is over the surface $\mathcal{S}$:
	\begin{equation}
	E\left( f \right) = \int_S {{{\left| {\nabla f\left( v \right)} \right|}^2}} dv = \int_S {f\left( v \right)\Delta f\left( v \right)} dv.
	\label{con:single_dirichlet}
	\end{equation}
	In its discrete form, the Dirichlet energy is computed as ${\pmb{\rm{f}}^{\rm{T}}}\mathbf{L}\pmb{\rm{f}}$. Combined with Equations \eqref{con:LB_eigen}, \eqref{con:WB}, \eqref{con:WFT} and \eqref{con:CONS},
	the Dirichlet energy of the function can be expressed in the graph wavelet basis as follows:
	\begin{align}
	{E}\left( \pmb{\rm{f}} \right) &= {\pmb{\rm{f}}^{\rm{T}}}{\bf{L}}\pmb{\rm{f}}
	= \sum\limits_{j{\rm{ = 0}}}^{N-1} {{\lambda _j}\left(\sum\limits_{m{\rm{ = 0}}}^{K}\sum\limits_v{\gamma _j\left( {t_m,v} \right)}\right)^2},
	\label{con:coeff_energy}
	\end{align}
	where $\gamma _j\left( {t_m,v} \right) =  {{W_{\pmb{\rm{f}}}}\left( {t_m,v} \right) g_{t_m}\left( {{\lambda _j}} \right){\phi _j}\left( v \right)}$ and $N$ is the number of vertices. We define $g_{t_m}\left( {{\lambda _j}}\right)$ as follows:
	\begin{equation}
	{g_{t_m}}\left( {{\lambda _j}} \right) = \left\{ {\begin{array}{*{20}{l}}
		{h\left( {{\lambda _j}} \right), \ \qquad if \quad m = 0}\\
		{g\left( {t_m{\lambda _j}} \right),\quad if \quad m >0  }
		\end{array}} \right.
	\end{equation}
	The Dirichlet energy of a vector-valued function $\pmb{\rm{F}}=\left(\pmb{\rm{f}}_1,\pmb{\rm{f}}_2,...,\pmb{\rm{f}}_{\rm{d}}\right): V\to \mathbb{R}^d$ on the mesh is defined as the sum of the Dirichlet energy of the individual components:
	\begin{equation}
	E\left( \bm{F} \right) = \sum\limits_{i = 1}^d{{E}\left( \pmb{\rm{f}}_i \right)}.
	\label{con:multiple_dirichlet}
	\end{equation}
	Now, we substitute the coordinate functions
	$\pmb{\rm{X}}=(\pmb{\rm{x}}_1, \pmb{\rm{x}}_2, \pmb{\rm{x}}_3): V\to \mathbb{R}^3$ into the Equation \eqref{con:multiple_dirichlet} above and express the result in the wavelet basis following Equation \eqref{con:coeff_energy}:
	\begin{align}
	{E}\left( \pmb{\rm{X}} \right)
	&= \sum\limits_{i = 1}^d {\sum\limits_{j{\rm{ = 0}}}^{{N-1} }{\lambda _j}\sum\limits_{m{\rm{ = 0}}}^{K}\sum\limits_v {\gamma_{ij}{\left( {t_m,v} \right)} {\omega _{ij}}} }  \\
	&= \sum\limits_{m{\rm{ = 0}}}^{K}\sum\limits_v \sum\limits_{j{\rm{ = 0}}}^{{N-1} } {{\lambda _j}\sum\limits_{i = 1}^d {\gamma_{ij}{\left( {t_m,v} \right)} {\omega _{ij}}} },
	\end{align}
	where ${\omega _{ij}} = \sum\limits_{m{\rm{ = 0}}}^{K}\sum\limits_v{\gamma _{ij}\left( {t_m,v} \right)}$.
	The Dirichlet energy of this generalized function can be decomposed into different scale energies $\sum\limits_{j{\rm{ = 0}}}^{{N-1} } {{\lambda _j}\sum\limits_{i = 1}^d {\gamma_{ij}{\left( {t_m,v} \right)} {\omega _{ij}}} }$ on each vertex.

	In addition, we ignore the first term when $j=0$ since the first eigenvalue of a mesh $\lambda_0 = 0$.
	The Dirichlet energy can be decomposed into $K$ multiscale vectors with lengths equal to the number of vertices and can be expressed as: 
	\begin{equation}
	\bm{\varepsilon}_{t_m} = \left\{ {\sum\limits_{j{\rm{ = 1}}}^{N - 1} {{\lambda _j}\sum\limits_{i = 1}^d {{\gamma _{ij}}\left( {t_m,v} \right){\omega _{ij}}} } } \right\}, m \in \left[ {0,{K}} \right].
	\label{energy_decomposition}
	\end{equation}
	
	\begin{figure}[!t]
		\centering
		\input{filter_good.te}\hspace{-0.4cm}
		\input{filter_bad.te}
		\begin{overpic}[trim=9cm -3cm 10cm 0cm, clip,width=0.4\linewidth, grid=false]{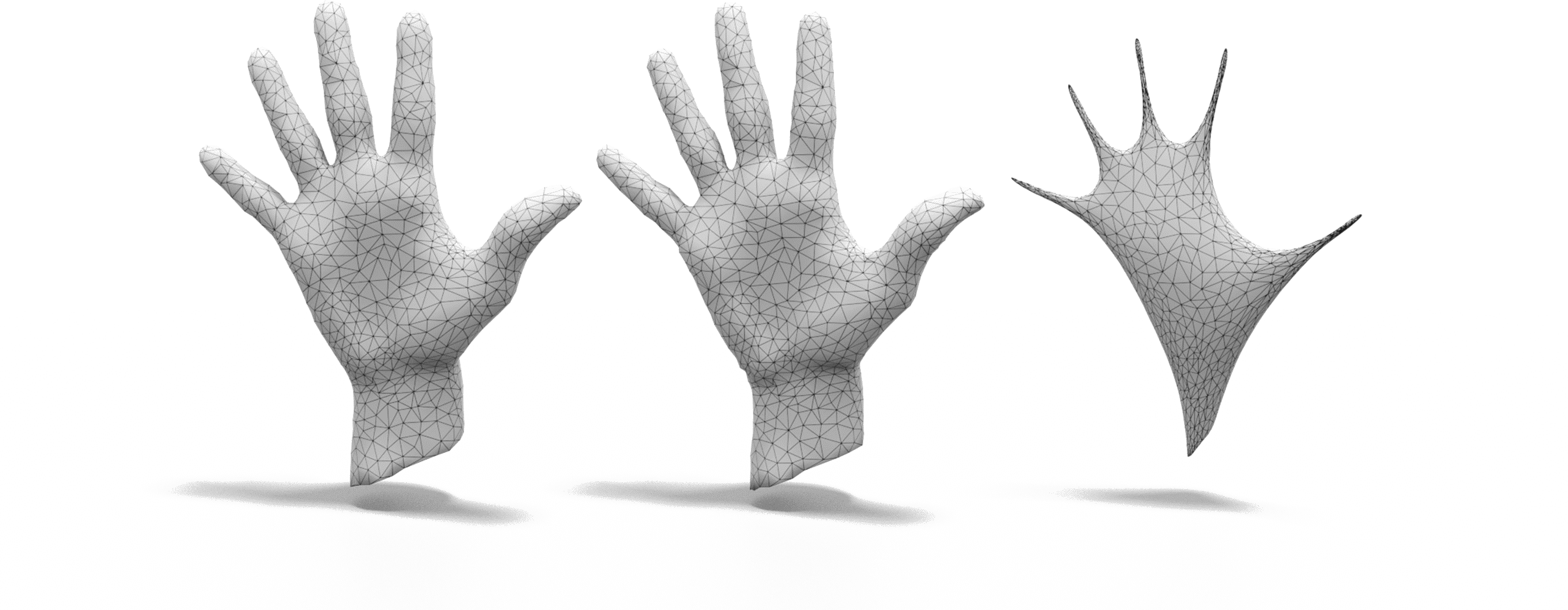}
			\put(5,64){\footnotesize Target}
			\put(40,64){\footnotesize Good}
			\put(75,64){\footnotesize Bad}
			\put(3,5){\footnotesize Use Equation~\eqref{con:CONS} for reconstruction}
		\end{overpic}\vspace{-12pt}
		\caption{ We show two choices of filters (dashed lines) and the function $G$ defined in Equation~\eqref{con:filters} (solid black line). We then use these two sets of filters to reconstruct a hand shape, \emph{i.e.}, use Equation~\eqref{con:CONS} to reconstruct the 3D coordinate functions. We can see that for the bad filter choice, where the constraint $G = 1$ is not satisfied, the reconstruction quality is very poor. On the other hand, the good filters can better reconstruct the given signal.}
		\label{fig:filters}\vspace{-12pt}
	\end{figure}
	
	After energy decomposition, the amount of energy of an individual vertex depends on the number of vertices on the surface. Because the Dirichlet energy of a shape with different resolutions or discretizations is constant, the higher the resolution of a mesh is, the less energy each vertex has. To get features of similar scale at a surface point when the underlaying discretization changes, we need to collect local energy at each vertex to form our signature. Finding the geodesic neighbors of a surface point is very time-consuming. Fig.~\ref{fig:wavelet_local} shows the wavelet basis functions at the meshes of different resolutions. We find that the shape of the wavelet does not change significantly on the meshes of different resolutions, which can be used to weight different resolutions of meshes. Thanks to the natural local properties of graph wavelets, the local wavelets can be used to collect the local energy to compute our signature on different resolutions. A wavelet $\bm{\psi} _{{t_s,v}}$ at a particular vertex can be thought of as the associated weights of points with the particular vertex. The weights of points are significant when the points near the specific vertex when measured by geodesic distance, and the influence is small if points is far from the vertex. For one scale $t_s$ and every vertex $v$, we first normalize the wavelet $\bm{\psi} _{{t_s,v}}$ to obtain a normalized vector $\bm{\psi} _{{{t_s},v}^ * }$ using minmax normalization.
	Then, we use normalized vector $\bm{\psi} _{{{t_s},v}^ *}$ to weight energy $\bm{\varepsilon}_{t_m}$. The formula of our signature on one vertex $v$ at one scale $t_s$ is expressed as follows:
	\begin{equation}
	WEDS_{t_s} \left( v \right)= \left\{ \sum\limits_x{\psi _{_{{t_s},v}}^ *\left( x \right)} {\varepsilon_{t_m} \left( x \right) }\right\}, m \in \left[ {0,{K}} \right].
	\label{spectral_feature}
	\end{equation}

	To obtain scale invariance, the energy vectors $\bm{\varepsilon}_t$ are modified by multiplying its eigenvalue $\lambda _j$ like $\rm{LPS}$~\cite{Wang_2019_CVPR}. Therefore, $\bm{\varepsilon}_{t_m} = \left\{\sum\limits_{j{\rm{ = 1}}}^{N - 1} {{\lambda _j^2}\sum\limits_{i = 1}^d {{\gamma _{ij}}\left( {t_m,v} \right){\omega _{ij}}} }  \right\}$.
	To construct our vertex descriptor, the weighting approach with one wavelet scale is not discriminative. Therefore, we cascade the descriptors at different scales in the scale set $\bm{S}_{t_s}$: $\text{WEDS} \left( v \right)= \left\{ \text{WEDS}_{t_s} \left( v \right)\right\}, t_s \in \bm{S}_{t_s}$. We will mention how to select scales in Section~\ref{sec:Multiscale_Filters}.

    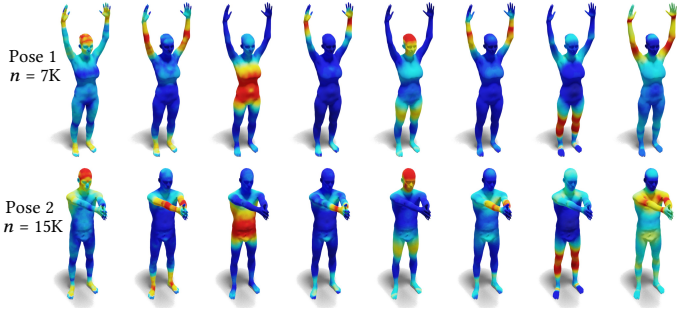
\begin{figure}[!t]
    	\begin{overpic}[trim=0cm 0.8cm 0cm -2cm,clip,width=0.98\linewidth,grid=false]{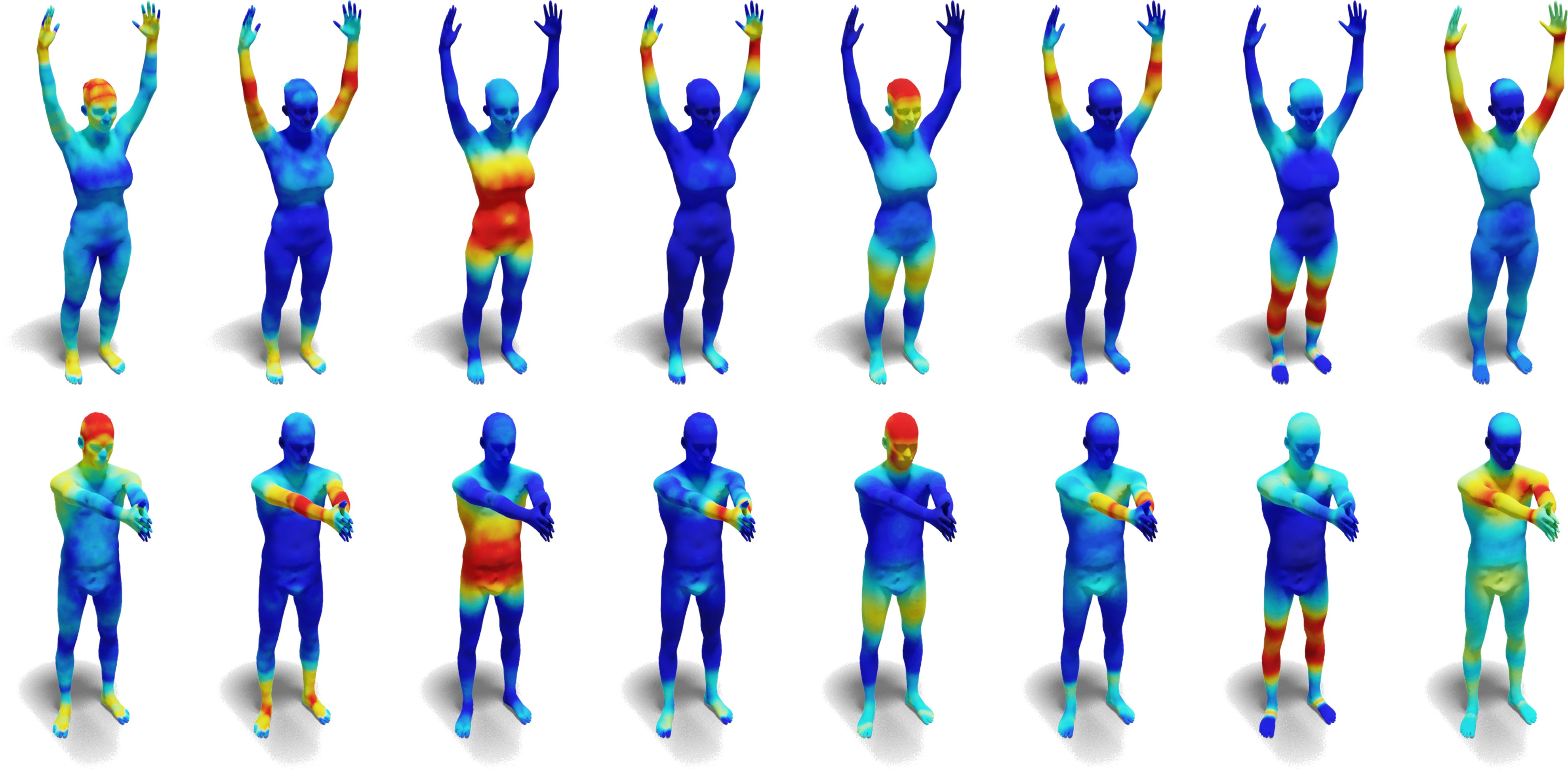}
    		\put(-6,39){\footnotesize Pose 1}
    		\put(-6,36){\footnotesize $n$ = 7K}
    		\put(-6.5,15){\footnotesize Pose 2}
    		\put(-7,12){\footnotesize $n$ = 15K}
    	\end{overpic}\vspace{-6pt}
    	\caption{We show WEDS on two different shapes with different poses and resolutions. We show 8 different dimensions of our descriptors. We can see that WEDS is robust with respect to the triangulation and resolution.}
    	\label{fig:eg:desc_weds}
    	\vspace{-12pt}
    \end{figure}
	
	\subsection{Multiscale Filters}
	\label{sec:Multiscale_Filters}
	
	If an original signal can be recovered by a graph wavelet basis (recall Equation \eqref{con:CON}), the filters need to satisfy the Parseval frame,
	\begin{equation}
	\label{con:filters}
	G(\lambda_j) = h^2\left( {{\lambda _j}} \right) + \sum\nolimits_{m{\rm{ = 1}}}^{K} g^2\left( {t_m{\lambda _j}} \right)  \equiv  1  ,
	\end{equation}
	and the proof is given in Appendix. Fig.~\ref{fig:filters} visualizes the importance of constraint $G$. We show two choices of filter functions, one that does and one that does not satisfy this constraint. We choose the Mexican hat functions as the filters of the graph wavelet, which are given by:
	$$
	g\left( {t_m{\lambda _j}} \right) = A{\left( {t_m{\lambda _j}} \right)^2}{e^{\left( {1 - {{\left( {t_m{\lambda _j}} \right)}^2}} \right)}},\quad
	h\left( {{\lambda _j}} \right) = B\,{e^{\left( { - {{\left( {\frac{{C{\lambda _j}}}{{{\lambda _{\max }}}}} \right)}^3}} \right)}},
	$$
	where $t_m = {e^{\text{linspace}\left( {\log \left( {\frac{{{D}}}{{{\lambda _{\max }}}}} \right),\log \left( {\frac{{{E}}}{{{\lambda _{\max }}}}} \right),K} \right)}}$. Considering the balance of efficiency and accuracy, in our tests, we choose 32 wavelet filters, \emph{i.e.}, $K = 31$. We set the tolerance to 0.01, and the five parameters can be solved: $A=0.443, B=1.004, C=38.462, D=46, E=0.2$.

	For the scale set $\bm{S}_{t_s}$, the selection is based on the number of dimensions generated by the features on each vertex. For one scale in $\bm{S}_{t_s}$, we can generate 32-dimensional features. So WEDS can generate features with a maximum of 1024 dimensions. But high-dimensional features are usually not needed. To represent high, medium, and low frequencies, we select at least 3 scales, and features with  at least 96 dimensions are generated, after which we can obtain fewer dimensions by sampling. If more dimensions of feature need to be generated, then more scales need to be picked. We take a linear sampling approach and remove the first and last filters, which is as follows:
	\begin{equation}
	\bm{S}_{t_s}= \left\{ {{t_m},m = \left\lfloor {{\rm{linspace(32,1,}}\left\lceil {{\rm{\frac{Num}{32}}}} \right\rceil {\rm{  +  2)}}} \right\rfloor \left( {{\rm{2:end - 1}}} \right)} \right\},
	\end{equation}
	where $\rm{Num}$ is the feature dimension of the output. In Fig.~\ref{fig:eg:desc_weds}, we show our WEDS descriptors on two shapes with 8 selected dimensions.

	\setlength{\tabcolsep}{0.9em}
	\begin{table}[!t]
		\caption{The discrete Dirichlet energy $E( \pmb{\rm{f}} )$ of the vertex coordinate functions and the WKS functions.}
		\label{tab:energy}
		\vspace{-8pt}
		\footnotesize
		\begin{tabular}{|c|c|c|c|c|c|c|}
			\hline
			\multirow{4}{*}{\shortstack{$E( \pmb{\rm{f}} )$ \\ of Coord}}  & Res & Dims1 & Dims2 & Dims3 & SUM & Area*2 \\ \cline{2-7}
			& 6890 & 1.0153 & 1.4768 & 1.1114 & 3.6035 & 3.6035 \\ \cline{2-7}
			& 10K & 1.0156 & 1.4716 & 1.1092 & 3.5964 & 3.5964 \\ \cline{2-7}
			& 15K & 1.0135 & 1.4719 & 1.1061 & 3.5915 & 3.5915 \\ \hline
			\multirow{4}{*}{\shortstack{$E( \pmb{\rm{f}} )$ \\ of WKS}} & Res & Dims1 & Dims2 & Dims3 & Dims4 & SUM \\ \cline{2-7}
			& 6890 & 0.7645 & 2.9629 & 4.8548 & 4.0352 & 12.6174 \\ \cline{2-7}
			& 10K & 0.7639 & 2.9764 & 4.8979 & 4.0372 & 12.6754 \\ \cline{2-7}
			& 15K & 0.7581 & 2.9836 & 4.7605 & 4.0883 & 12.5905 \\ \hline
		\end{tabular}
	    \vspace{-30pt}
	    \caption*{}
	    \vspace{-6pt}
	\end{table}
	
	\subsection{Discussion}       
	\label{subsec:discussionofWEDS}
	
	Many spatial and spectral descriptors have been proposed, but these descriptors cannot satisfy the expected property simultaneously, such as resolution, scale, and discrimination.
	Our goal is to find a new descriptor that can be discriminative and robust to different shape structure at the same time. We are inspired by the following observation: for a smooth surface $\mathcal{S}$ to any triangulated mesh $\mathcal{M}$ with any resolution. If a discrete vector $\pmb{\rm{f}}$ is sampled from a smooth function $f$, the discrete Dirichlet energy $E( \pmb{\rm{f}} ) =
	{\pmb{\rm{f}}^{\rm{T}}}\mathbf{L}\pmb{\rm{f}} $ is robust to discretization. Table~\ref{tab:energy} shows two smooth functions, which is vertex coordinate and WKS.
	
	It can be found that the discrete Dirichlet energy on every dimension and its sum on this two smooth functions are robust to the change of resolution. In addition, Dirichlet energy is invariant to a rigid transformation, which is very important in feature design. Therefore, we want to derive a set of descriptors from the Dirichilet energy of a given function $\pmb{\rm{f}}$.
	An interesting phenomenon is that the Dirichlet energy of vertex coordinates has been proved to be twice of surface area of the mesh, and the surface area is very robust to different discretization. The coordinates are also the most primitive and comprehensive information of given shape, so we choose the vertex coordinate function as input. To have other desirable properties, we just need to pick a different function $\pmb{\rm{f}}$.
	
	To derive a set of per-vertex descriptors from this energy, we need to distribute the energy of the function $\pmb{\rm{f}}$ to the vertices. One trivial solution is simply use the Laplacian-Beltrami basis, where we can project function $\pmb{\rm{f}}$ to the Laplacian-Beltrami basis like Equation \eqref{con:coeff}. However, in this case, we only have ${\sigma _j}$ that characterizes the global attributes of the shape. One possible way to get local features for each vertex is to cut a geodesic disk like LPS~\cite{Wang_2019_CVPR}, and compute the Dirichlet energy locally, but it is time-consuming and does not contain global information.
	
	Another choice is to distribute the energy of $\pmb{\rm{f}}$ using the graph wavelet basis, which are a set of basis defined on each of the vertices. In this case, the wavelet basis can capture the local details in a spatial region around the vertex. In addition, wavelets on graphs are expressed using eigenfuctions of the Laplacian-Beltrami operator, and we can project function $\pmb{\rm{f}}$ to the wavelet basis and get the coefficients in Equation \eqref{con:WFT}. After projection, it can be found that the wavelet coefficient $W_{\pmb{\rm{f}}}(t,v)$ contains the coefficient ${\sigma _j}$ which includes global information of $\pmb{\rm{f}}$. Because the coefficients of graph wavelets can capture both global and local information, the reconstructed energy of $\pmb{\rm{f}}$ can be distributed to each vertex while maintaining the global and local information at the same time. Therefore, we derive a descriptor with high discrimination while maintaining robustness.
	To sum up, wavelets enable us to achieve a trade-off between local and global information that other descriptors are unable to achieve.

	\section{Supervised Descriptor Learning}  
	
	We propose a new graph convolutional network called MGCN. We mainly employ MGCN for descriptor learning in this paper, but it is a general architecture for graph convolutional networks. We first describe a single layer of our network in Section~\ref{sec:Layer}, and then the complete architecture and training details in Section~\ref{sec:Architecture}.
	
	\subsection{Multiscale Graph Convolution Layer}       
	\label{sec:Layer}
	
	A layer of our network takes a $C$-dimensional vector for each of the $N$ graph nodes as input and outputs an $O$-dimensional vector for each node. For simplicity of notation, we start the description by considering the case of $C=O=1$ and extend to higher dimensions in the end. We focus on discrete descriptions for simplicity. Under this simplifying assumption, we denote the input as signal $\pmb{\rm{x}}^{\text{in}} \in {\mathbb{R}^N}$ and the output as $\pmb{\rm{x}}^{\text{out}} \in {\mathbb{R}^N}$. The goal of our layer is to convolve the signal $\pmb{\rm{x}}$ with a filter $\pmb{\rm{y}} \in {\mathbb{R}^N}$, \emph{i.e.}, to compute $\pmb{\rm{y}} {*_w} \pmb{\rm{x}}$ where $*_w$ is the convolution operator.
	Convolution in the time domain is equal to the product in the frequency domain. Following previous work, we consider spectral convolutions on graphs defined as:
	\begin{equation}
	\pmb{\rm{x}}^{\text{out}}=\pmb{\rm{y}} {*_w} \pmb{\rm{x}}^{\text{in}} = \bm{\Phi}\left(\left(\bm{{\Phi}^T}\pmb{\rm{y}}\right) \odot \left(\bm{{\Phi}^T}\pmb{\rm{x}}^{\text{in}}\right)\right) = \bm{\Phi}\pmb{\rm{{w}} }_\theta\bm{{\Phi}^T}\pmb{\rm{x}}^{\text{in}},
	\end{equation}
	where $\odot$ is the element-wise product, $\bm{\Phi} \in {\mathbb{R}^{N \times k}}$ is an eigenvector matrix, and $\pmb{\rm{w}}_\theta \in \mathbb{R}^{k \times k}$ is a diagonal matrix describing the filter in the frequency domain.
	In general, $\pmb{\rm{w}}_\theta$ can be considered to be a function of the eigenvalue matrix $\bm{\Lambda}$, \emph{i.e.},
	$\pmb{\rm{w}}_\theta = f(\bm{\Lambda})$.
	To improve efficiency, ChebyNet~\cite{defferrard2016convolutional} approximates the arbitrary function $f$ as weighted sum of powers of $\bm{\Lambda}$:
	\begin{equation}
	\pmb{\rm{w}}_\theta = \sum\limits_{m = 0}^{K-1} {{\theta _m}} \text{diag}{\left( {\left\{ {{\lambda _j}} \right\}_{j=0}^{k-1}} \right)^m} = \sum\limits_{m = 0}^{K-1} {{\theta _m}} \bm{{\Lambda} ^m},
	\end{equation}
	where $\bm{\Lambda}$ is a diagonal matrix of size $K \times K$, and the elements on the diagonal are eigenvalues $\lambda _j$.
	
	In this way, the convolution can be replaced by a linear combination of $m$-order polynomials of the Laplacian matrix, which eliminates the need to calculate the eigenfunctions. Laplacians with order $m$ are $m$-localized. For efficient computation, Chebyshev polynomials are computed recursively as follows:
	\begin{equation}
	\pmb{\rm{x}}^{\text{out}} = \pmb{\rm{y}} {*_w} \pmb{\rm{x}}^{\text{in}} \approx \sum\limits_{m = 0}^K {\theta _m} {T_m}\left( \bf{L}  \right)\pmb{\rm{x}}^{\text{in}},
	\end{equation}
	where ${T_m}\left( \bm{{L} } \right) \in {\mathbb{R}^{N \times N}}$ is the $m$-order Chebyshev polynomial.
	
	From another point of view, the convolution of the ChebyNet can be understood as the sum of polynomials of different orders evaluated at the Laplacian. If the order is $m$, the receptive field of the convolution is the $m$-ring neighborhood of a vertex.
	The ChebyNet is equivalent to the weighted sum of multiple convolutions with increasing receptive field.
	The problem of this convolution is that it is not resolution independent, because the size of the $m$-ring neighborhood depends on the discretization of the surface.
	
	Our idea is to express $\pmb{\rm{w}}_\theta$ in a wavelet filter basis, rather than a polynomial basis. Note that there is a difference between the wavelet filter basis and the wavelet basis. The wavelet filter basis is a basis in the spectral domain, but the wavelet basis exists in the spatial domain on the surface:
	\begin{equation}
	\pmb{\rm{w}}_\theta  = \sum\limits_{m = 0}^{K} {{\theta _m}} \text{diag}{\left( {\left\{ {{g_{t_m}\left( {{\lambda _j}} \right)}} \right\}_{j=0}^{k}} \right)} = \sum\limits_{m = 0}^{K} {{\theta _m}} { g_{t_m}\left( \bm{{\Lambda}} \right)}.
	\end{equation}
	Now the convolution can be simplified as follows:
	\begin{equation}
	\pmb{\rm{x}}^{\text{out}} = \pmb{\rm{y}} {*_w} \pmb{\rm{x}}^{\text{in}} \mathop  \approx \limits \sum\limits_{m = 0}^{K} {{\theta _{m}}} \bm{\Psi} _{{t_m}}^T \pmb{\rm{x}}^{\text{in}},
	\end{equation}
	where each matrix $\bm{\Psi} _{{t_m}} \in \mathbb{R} ^{N \times N}$ is composed of the wavelet basis with scale $t_m$.

	In practice, there are three factors to consider.
	First, the magnitude of high frequency wavelet functions is very small, which leads to numerical robustness issues during learning the parameters.
	Second, we need to ensure that the convolution operation leads to similar results for surfaces that are discretized with different sampling densities.
	Therefore, to solve the first two issues, we perform $L_1$ normalization on the wavelet functions (\emph{i.e}., normalize the columns of the wavelet matrix $\bm{\Psi})$.
	Finally, we do not need to select all the filters, we only need to select a part (such as sampling about half) of the filters to reduce the calculation. Therefore, the new convolution is as follows,
	\begin{equation}
	\pmb{\rm{x}}^{\text{out}} = \pmb{\rm{y}}{*_w}\pmb{\rm{x}}^{\text{in}} \approx \sum\limits_{{t_s} \in {\bm{S}_{{t_s}}}} {{\theta _{{t_s}}}} { {{\overline{\bm{\Psi} _{{t_s}}}}} ^T}\pmb{\rm{x}}^{\text{in}},
	\end{equation}
	where $\overline{\bm{\Psi} _{{t_s}}}$ is a normalized matrix with each wavelet $L_1$ normalized, where the sum of the normalized wavelet is 1.
	
	\begin{figure}[!t]
		\centering
		\includegraphics[width=1\linewidth]{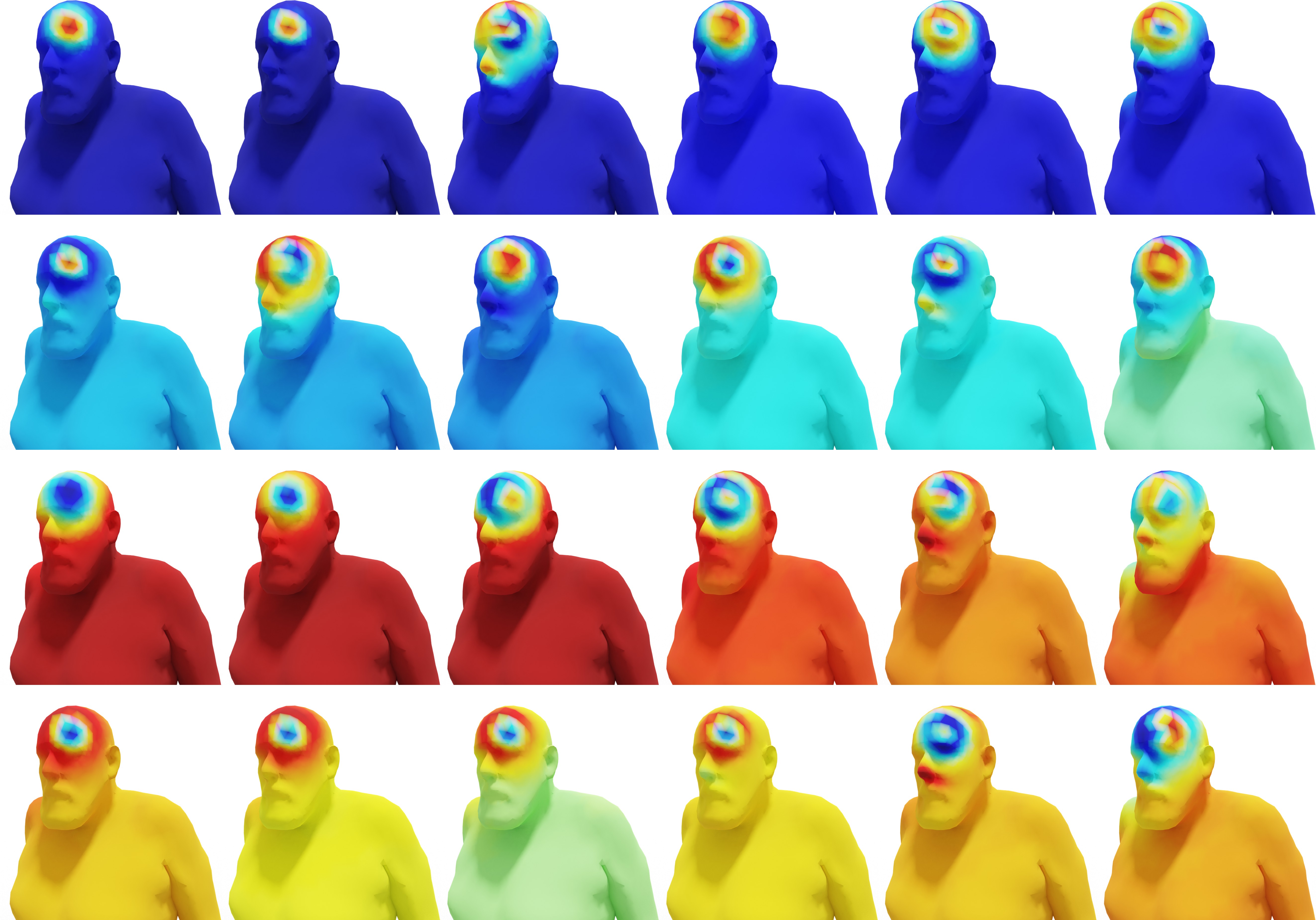}\vspace{-6pt}
		\caption{Visualization of filters learned by our network. For one input mesh and one selected vertex on the head, we show 24 manually selected and normalized filters to ensure some variability. Note that the network learns about 60K filters so this is only a small subset. Further, the filters are normalized, since the range of values between different filters differs significantly.}\vspace{-9pt}
		\label{fig:filter_visualization}
	\end{figure}
	
	For the high-dimensional case with $\bf{X}^{\text{in}} \in {\mathbb{R}^{N \times C}}$, we form our convolution module by the following formula:
	\begin{equation}
	\bf{Z} = {\rm{Norm}}\left( {{\rm{ELU}}\left( {\sum\limits_{{t_s} \in {\bm{S}_{t_s}}} {{{\overline {\bm{\Psi} _{{t_s}}} }^T}\bf{X}^{\text{in}}{\bf{W}_{{t_s}}}} } \right)} \right){\rm{ }},
	\end{equation}
	where ${\bf{W}_{{t_s}}} \in {\mathbb{R}^{C \times O}}$ and $O$ are the dimensions of the output feature, and Norm() is a minmax normalization on every dimensional feature to normalize energies of vertices. This module can be called as ``MGCONV''.

	\subsection{Network Architecture Details}      
	\label{sec:Architecture}
	
	The MGCONV layer described previously can be used to construct our MGCN. Here, we describe the architecure we used for desriptor learning.
	To learn a shape descriptor, we build an MGCN network by stacking 6 layers of MGCONV and one fully connected layer: $5 \times$ MGCONV96(16) + MGCONV128(16) + FC256.
	MGCONVx(k) refers to a convolutional layer that has an x-dimensional output of feature maps, and (k) refers to scale $k$ of our wavelet scale set.
	and FCx refers to a fully connected layer that outputs a vector with x-dimension, and 256 refers to the 256 dimension of the output feature. \revised{To have a fair comparison and equal number of parameters in the network, the dimension of all input features is set to 128.}
	
	\revised{We use two phases of training. In the first phase, a classification network is used to train the MGCN. One fully connected layer FCd is added after the last MGCONV layer, and the cross-entropy loss is used to classify each point.
		For the second phase, we propose to use the HardNet loss~\cite{mishchuk2017working}. Its advantage is that it can directly calculate the loss of N pairs by computing N triplet distances. We apply this loss to the task of learning shape descriptors. The HardNet loss is used to directly train MGCN to reduce the distance between positive examples and increase the distance between negative examples. During training, we use a batch size of 1 because the shapes in the dataset can have a different number of vertices. In our tests, we first use training with cross-entropy loss for 200 epochs and then train with HardNet loss for 100 epochs. Compared with training the network only using cross-entropy loss with 300 epoch, this configuration can further reduce average geodesic error about 0.01. Training using only the HardNet loss is slower, so the first phase can be considered an acceleration strategy to have a good initialization.
		For training with the cross-entropy loss, we use the ADAM optimizer (${\beta _1} = 0.9, {\beta _2} = 0.999, \varepsilon  = {10^{ - 8}}$) with a learning rate of ${10^{ - 3}}$ and a weight decay of ${10^{ - 4}}$.  For training with the HardNet loss, we use the ADAM optimizer with a learning rate of ${5 \times 10^{ - 4}}$ and a weight decay of ${5 \times 10^{ - 5}}$. }
	
	To give some intuition of the learned filters, we visualize a small subset of them in Fig.~\ref{fig:filter_visualization} for a training on the FAUST dataset.
	
	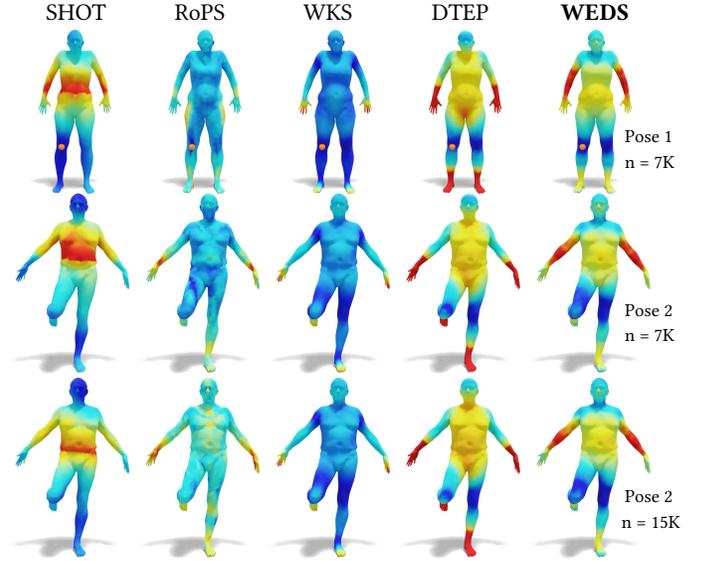
\begin{figure}[!t]
		\centering
		\begin{overpic}[trim=0cm 0.8cm 0cm -1cm,clip,width=1\linewidth,grid=false]{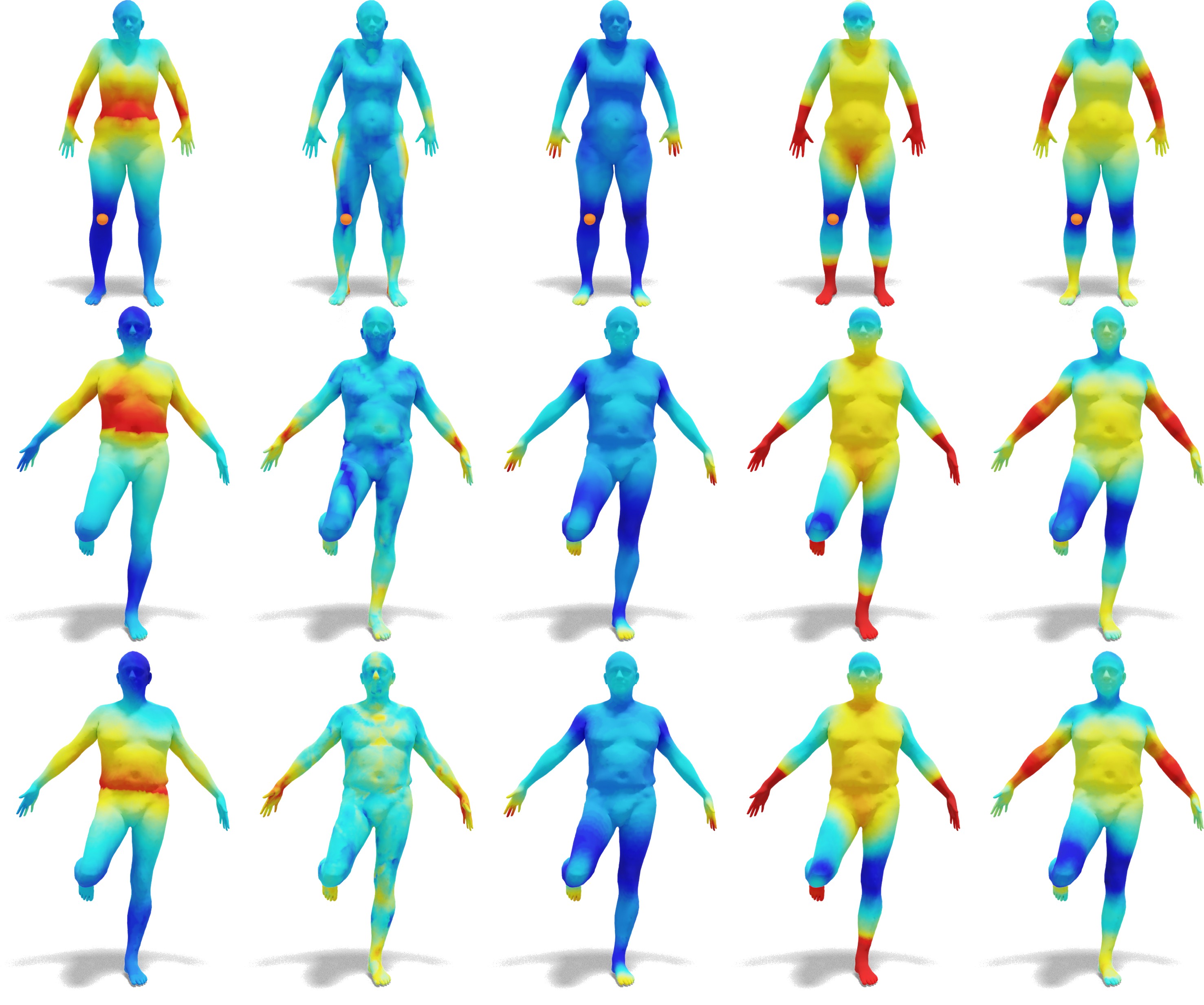}
			\put(6,84){SHOT}
			\put(26,84){RoPS}
			\put(46,84){WKS}
			\put(66,84){DTEP}
			\put(86,84){\textbf{WEDS}}
			\put(96,65){\footnotesize Pose 1}
			\put(96,61){\footnotesize n = 7K}
			\put(96,38){\footnotesize Pose 2}
			\put(96,34){\footnotesize n = 7K}
			\put(96,9){\footnotesize Pose 2}
			\put(95.5,5){\footnotesize n = 15K}
		\end{overpic}
		\vspace{-20pt}
		\caption{For a selected vertex on the knee (shown in orange) we visualize the  distance of the descriptor on the selected vertex to the descriptors of other vertices. A blue color indicates a small distance and a red color a large distance. Top row: descriptor distances on the same mesh. Second row: descriptor distances on a different mesh in the same resolution. Third row: descriptor distances on a different mesh in a different resolution. Form left to right: SHOT, RoPS, WKS, DTEP, and WEDS. We can observe that WEDS and DTEP are more discriminative than WKS and that SHOT and RoPS are not resolution independent.}
		\label{fig:distance_vis}\vspace{-12pt}
	\end{figure}
	
	\subsection{Discussion}
	Spectral CNN exploits the fact that the convolution in the time domain is equal to the product in the frequency domain, but this network leads to descriptors that are too smooth and do not contain enough local information. ChebyGCN uses $k$-order Chebyshev polynomials to expand filters in the frequency domain (Note that in this section $k$ does not refer to the size of the Laplacian basis). GCN further simplifies ChebyGCN and uses a $1$-ring neighborhood to approximate filters to reduce the amount of calculations. By contrast, SplineCNN and DGCNN belong to the class of networks using spatial convolution. The core questions of the spatial convolution method are how to find neighbors and how to aggregate the information from the neighbors. Comparing convolution in the frequency domain to the convolution in the spatial domain, it can be seen that the locality is important to improve performance. But this locality also brings some problems. For example, GCN and SplineCNN use $1$-ring neighborhood information. ChebyGCN uses $k$-order Laplacian polynomials, so it uses $k$-ring neighborhood information. DGCNN uses $k$-nearest neighbors in feature space to define the support of a convolution filter. However, such convolution operations do not generalize well to meshes of different resolution, because the size of a $k$- or $1$-ring neighborhood (the receptive field of the filters) changes with the resolution of a mesh.
	An alternative method is to calculate a geodesic disk centered at each vertex and locally resample the surface~\cite{Wang_2019_CVPR}. However, computing dense geodesic disk is time-consuming and resampling the surface introduces additional errors.
	
	An alternative approach is the graph wavelet neural network~\cite{XuSCQC19}. The formula for convolution is $\pmb{\rm{y }} {*_\mathbf{g}} \pmb{\rm{x}} = \bm{\Psi} \mathbf{g}_\theta \bm{{\Psi} }^{{ - 1}}\pmb{\rm{x}}^{\text{in}}$. However, because the filter $\mathbf{g}_\theta$ has parameters equal to the number of vertices, the network only works for meshes with the same number of vertices. Second, the calculation of matrix inversion is very slow for large graphs. In summary, there is currently no reliable network to handle multi-resolution datasets.
	
	In our MGCONV layer, we use many advantages of the nature of wavelets, but make important changes to the existing graph wavelet neural network. First, the spectral filters can be expressed in the wavelet filter basis. In the spatial domain, convolution operations are replaced by multiplying with a matrix composed of the wavelet basis. Because the wavelet basis functions are robust to the change of resolution and triangulation, our convolution also has the ability to cope with different resolutions. Second, due to the multiscale nature of the wavelet basis, we have the ability to simultaneously capture local and global information. Most importantly, we can capture local information without the explicit computation of local neighborhoods using geodesic distances as in~\cite{Wang_2019_CVPR}.

	
	\section{Experimental Results}  
	
	We present multiple qualitative results. After describing the used evaluation metrics, we compare WEDS with other \revised{non-learned} descriptors, compare MGCN with other network architectures for descriptor learning, and evaluate different parameter settings.
	In our experiments, the results are obtained using an Intel Core i7-7700 processor with 4.2 GHz and 16 GB RAM. Offline training is run on an NVIDIA GeForce GTX RTX (24 GB memory) GPU.

	\begin{figure}[!t]
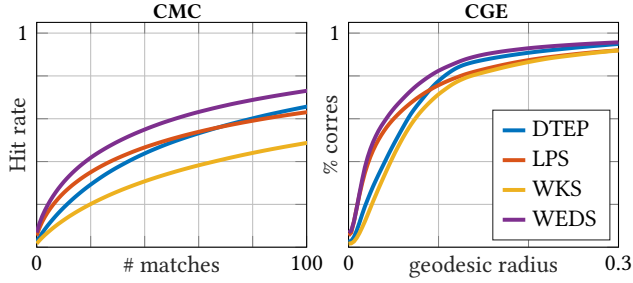
\vspace{-6pt}
		\centering
		\input{curve_faust_nonlearned_symm_cmc.te}\hspace{-15pt}
		\input{curve_faust_nonlearned_symm_kim.te}\vspace{-10pt}
		\caption{The symmetric CMC and CGE metrics of \revised{non-learned} descriptors on the FAUST dataset. We use different \emph{line colors} to indicate different networks. Note that our WEDS descriptor is the most discrimintive especially according to the CMC curves.}
		\label{fig:unsupervised_curce_faust}\vspace{-12pt}
	\end{figure}
	
	\begin{figure}[!t]
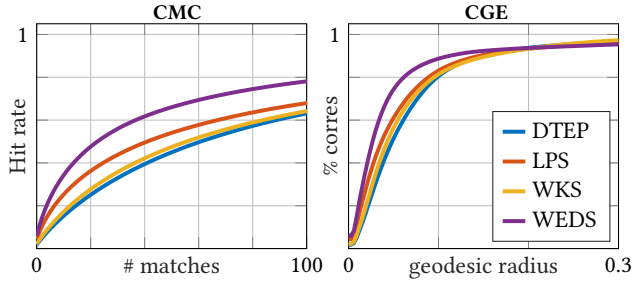

		\centering
		\input{curve_scape_nonlearned_symm_cmc.te}\hspace{-15pt}
		\input{curve_scape_nonlearned_symm_kim.te}\vspace{-10pt}
		\caption{The symmetric CMC and CGE metrics of \revised{non-learned} descriptors on the SCAPE dataset. We use different \emph{line colors} to indicate different networks. Note that our WEDS descriptor is the most discrimintive especially according to the CMC curves.}
		\label{fig:unsupervised_curce_scape}\vspace{-8pt}
	\end{figure}
	
	\subsection{Evaluation Metrics}
	We use three metrics to evaluate descriptors: average geodesic error, cumulative geodesic error, and cumulative match characteristic. We report results for two types of ground truth matches: the closest vertex on the direct map and the closest vertex on the symmetric map. Specifically,
	\begin{enumerate}[leftmargin=0cm,itemindent=.5cm,labelwidth=\itemindent,labelsep=0cm,align=left]
		\item Average geodesic error is a scalar to compute the average per-vertex error. The direct error evaluates the geodesic error between a predicted vertex and its ground-truth correspondence. The symmetry-aware error is the minimum of the direct error and the error between predicted vertex and the symmetric ground truth. The predicted vertex is obtained by computing the nearest-neighbor using the $L_2$ distance in feature space. \revised{For simplicity, the errors reported in all tables are scaled by $10^{-3}$.}
		\item Cumulative geodesic error (CGE) measures matching quality by plotting the percentage of nearest-neighbor correspondences that are at most $r$-geodesically distant from the ground-truth correspondence. According to the type of ground truth used, CGE is also divided into direct CGE and symmetry-aware CGE.
		\item Cumulative match characteristic (CMC) evaluates the percentage of vertices that find a correct match among the $k$-nearest neighbors in descriptor space. According to the type of ground truth used, CMC is also divided into direct CMC and symmetry-aware CMC.
	\end{enumerate}

	For the network evaluation, we use the descriptor evaluation if the network is used for learning descriptors.
	
	\subsection{Non-learned Descriptor Evaluation}
	\label{sec:Non-learned}
	As competitors, we select three spatial domain descriptors (SI~\cite{johnson1999using}, SHOT~\cite{tombari2010unique}, RoPS~\cite{guo2013rotational}) and four spectral domain descriptors (HKS~\cite{sun2009concise}, WKS~\cite{aubry2011wave}, LPS~\cite{Wang_2019_CVPR}, and DTEP~\cite{melzi2018discrete}).

	\setlength{\tabcolsep}{2.4em}
	\begin{table}[!t]
		\caption{Average (direct/symmetry-aware) geodesic error computed on 15$\times$14 shape pairs with different descriptors. WEDS improves about 10\% compared to the best competitor.}\vspace{-6pt}
		\label{tab:non-learned_acc}
		\footnotesize
		\begin{tabular}{|c|c|c|}
			\hline
			\multirow{2}{*}{Descriptors} & \multicolumn{2}{c|}{Dataset} \\ \cline{2-3}
			& FAUST(6890) & SCAPE(12.5K) \\ \hline
			SI & 352 / 153 & 380 / 251 \\ \hline
			SHOT & 381 / 262 & 271 / 196 \\ \hline
			RoPS & 346 / 252 & 267 / 187 \\ \hline
			HKS & 511 / 396 & 507 / 409 \\ \hline
			WKS & 335 / 118 & 260 / 72 \\ \hline
			LPS & 325 / 97 & 227 / 71 \\ \hline
			DTEP & 312 / 89 & 267 / 76 \\ \hline
			WEDS & \textbf{287 / 69}& \textbf{225 / 66} \\ \hline
		\end{tabular}\vspace{-3pt}
	\end{table}
	\setlength{\tabcolsep}{1.6em}
	\begin{table}[!t]
		\caption{\revised{Average direct geodesic error computed on 8$\times$7 non-isometric shape pairs with different descriptors.}}\vspace{-6pt}
		\label{tab:non-learned_SMAL}
		\footnotesize
		\begin{tabular}{|c|c|c|c|c|}
			\hline
			Descriptors & SI & SHOT & RoPS & HKS \\ \hline
			Error & 482 & 456 & 413 & 475 \\ \hline\hline
			Descriptors & WKS & LPS & DTEP & \textbf{WEDS} \\ \hline
			Error & 558 & 395 & 328 & \textbf{314} \\ \hline
		\end{tabular}\vspace{-6pt}
	\end{table}
	
	\subsubsection{Evaluation on near-isometric shapes}
	We first analyze the discriminative power of descriptors. To verify the effectiveness of the proposed local descriptor, we choose FAUST~\cite{Bogo2014faust} and SCAPE~\cite{Anguelov2005scape} as our test datasets. We also use an extended version of FAUST~\cite{Wang_2019_CVPR}, which contains meshes with different resolution, triangulation, scale, and rotation.

	We conduct an extensive evaluation for different \revised{non-learned} descriptors on FAUST and SCAPE. Between different humans, the pairs are isometric or \revised{near-isometric}.
	To reduce variability for fair test, we selected 15 models on two datasets randomly and test every two pairs of fifteen models. The total number of tested pairs is 15$\times$14. Table~\ref{tab:non-learned_acc} shows the average geodesic error A/B on 15$\times$14 pairs. A is direct error and B is symmetry-aware error.
	
	From Table~\ref{tab:non-learned_acc}, we find that RoPS is the better among the spatial domain descriptors, but the geodesic error is still large. The results show that spatial domain descriptors cannot handle non-rigid matching well. In addition to HKS being too smooth, frequency domain descriptors have better performance. Among the frequency domain descriptors, WEDS has the best discrimination  among state-of-the-art descriptors two datasets and has about 10\% performance improvement. A visualization of the distance of the descriptor of a selected vertex to other vertices' descriptors is shown in Fig.~\ref{fig:distance_vis}.  More details in the form of curves are provided in Fig.~\ref{fig:unsupervised_curce_faust} and Fig.~\ref{fig:unsupervised_curce_scape}. Here we see that the improvement of the CMC metric is even more significant than the average geodesic error. Other tests are given in additional materials.
	
		\setlength{\tabcolsep}{0.9em}
	\begin{table}[!t]
		\caption{Average (direct/symmetry-aware) geodesic error computed on 15$\times$14 shape pairs of FAUST with different descriptors. WEDS + MGCN significantly improves upon the best previous work LPS + Geodesic-based.}\vspace{-3pt}
		\label{tab:learn_faust}
		\footnotesize
		\begin{tabular}{|c|c|c|c|c|}
			\hline
			\multirow{2}{*}{Descriptors} & \multirow{2}{*}{Network} & \multicolumn{3}{c|}{\#Resolution} \\ \cline{3-5}
			&  & 6890 - 6890 & 6890 - 10K & 6890 - 15K \\ \hline
			Histogram & CGF & 424 / 298 & 433 / 300 & 460 / 332 \\ \hline
			- & OSD & 398 / 182 & 457 / 254 & 430 / 221 \\ \hline
			- & SplineCNN & 276 / 161 & 488 / 378 & 524 / 438 \\ \hline
			WEDS & ChebyGCN & \textbf{ 6 / 1} & 527 / 387 & 551 / 377 \\ \hline
			WKS & Geo-based & 204 / 35 & 221 / 52 & 252 / 69 \\ \hline
			LPS & Geo-based & 164 / 22 & 203 / 43 & 223 / 55 \\ \hline
			WEDS & Geo-based & 147 / 24 & 207 / 34 & 239 / 39 \\ \hline
			WKS & MGCN & 19 / 13 & 88 / 34 & 124 / 69 \\ \hline
			LPS & MGCN & 18 / 12 & 44 / 24 & 84 / 39 \\ \hline
			WEDS & MGCN & 8 / 7 & \textbf{26 / 18} & \textbf{66 / 34} \\ \hline
		\end{tabular}\vspace{-9pt}
	\end{table}
	
	\begin{figure}[!t]
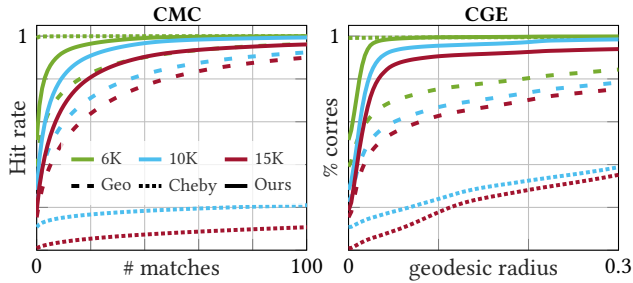

		\centering
		\input{curve_faust_direct_cmc.te}\hspace{-15pt}
		\input{curve_faust_direct_kim.te}\vspace{-9pt}
		\caption{The direct CMC and CGE metrics of learned descriptors on the FAUST dataset. We use different \emph{line types} to indicate different networks, where the Geo-based network~\cite{Wang_2019_CVPR} is in dashed lines, the ChebyCNN is in dotted lines, and our MGCN is in solid lines. We use different \emph{colors} to indicate different resolutions on the target shapes. Note that other networks do not generalize to different resolution as well as our network.}
		\label{fig:faust_learn}\vspace{-12pt}
	\end{figure}
	
	\subsubsection{Evaluation on non-isometric shapes}
	
	\revised{Non-isometric shapes are more challenging to match than near-isometric shapes. We choose SMAL~\cite{Zuffi2017} as our test dataset to compare different non-learned descriptors. SMAL is a small dataset with four-legged shapes such as lions, horses, cows, and hippos, where all base shapes share the same triangulation connectivity. A shape pair is considered non-isometric if two shapes are from different categories. We introduce the remeshed SMAL 5K dataset where all shapes are remeshed independently to test the effectiveness of different descriptors.
		
		For non-isometric shapes, we select 8 non-isometric models and test direct error on 8$\times$7 pairs. Table~\ref{tab:non-learned_SMAL} shows the results for different descriptors. From Table~\ref{tab:non-learned_SMAL}, we find that the descriptors directly constructed in the spatial domain such as SI, SHOT, and RoPS have average performance compared to other methods on non-isometric shapes. The intrinsic descriptors such as WKS and HKS have the worst performance. DTEP and WEDS are better than the other descriptors, but we believe that there is a lot of room for improvement.}

	\subsection{Graph Network Evaluation}

	To test the effectiveness of the network, we first test the shape descriptors that have been further improved by the network.
	The network structure has been described in Section~\ref{sec:Architecture}.
	
	\setlength{\tabcolsep}{1.0em}
	\begin{table}[!t]
		\caption{Average (direct/symmetry-aware) geodesic error computed on 10$\times$9 shape pairs of SCAPE with different descriptors. MGCN outperforms its best competitors ChebyGCN and Geo-based by a large margin.}\vspace{-3pt}
		\label{tab:learn_scape}
		\footnotesize
		\begin{tabular}{|c|c|c|c|}
			\hline
			\multirow{2}{*}{Descriptors} & \multirow{2}{*}{Network} & \multicolumn{2}{c|}{\#Resolution} \\ \cline{3-4}
			&  & SCAPE 5K & SCAPE 5K-12.5K(Test) \\ \hline
			Histogram & CGF & 374 / 264 & 428 / 323 \\ \hline
			- & OSD & 259 / 94 & 835 / 742 \\ \hline
			- & SplineCNN & 297 / 180 & 503 / 353 \\ \hline
			WEDS & ChebyGCN & 68 / 29 & 458 /  332 \\ \hline
			WKS & Geo-based & 185 / 72 & 192 / 80 \\ \hline
			LPS & Geo-based & 175 /  68 & 182 / 74 \\ \hline
			WEDS & Geo-based & 163 / 65 & 179 / 71 \\ \hline
			WKS & MGCN & 67 / 46 & 98 / 64 \\ \hline
			LPS & MGCN & 54 / 26 & 82 / 44 \\ \hline
			WEDS & MGCN & \textbf{48 / 17} & \textbf{73 / 39} \\ \hline
		\end{tabular}\vspace{-9pt}
	\end{table}

	\begin{figure}[!t]
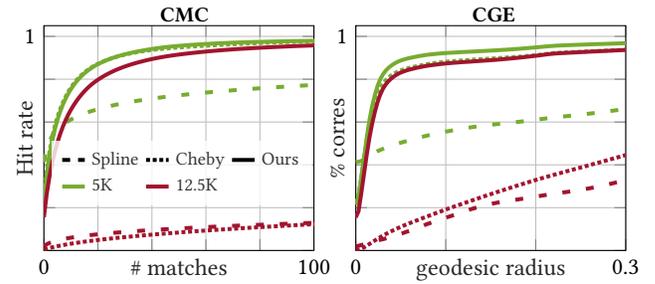

		\centering
		\input{curve_scape_direct_cmc.te}\hspace{-15pt}
		\input{curve_scape_direct_kim.te}\vspace{-9pt}
		\caption{The direct CMC and CGE metrics of learned descriptors on the SCAPE dataset. We use different \emph{line types} to indicate different networks and different \emph{colors} to indicate different resolutions on the target shapes. Note that other networks do not generalize to different resolution as well as our network.}\vspace{-12pt}
		\label{fig:scape_learn}
	\end{figure}

	\begin{figure*}[!t]
		\begin{overpic}[trim=0cm 0.8cm 0cm -2cm,clip,width=0.48\linewidth,grid=false]{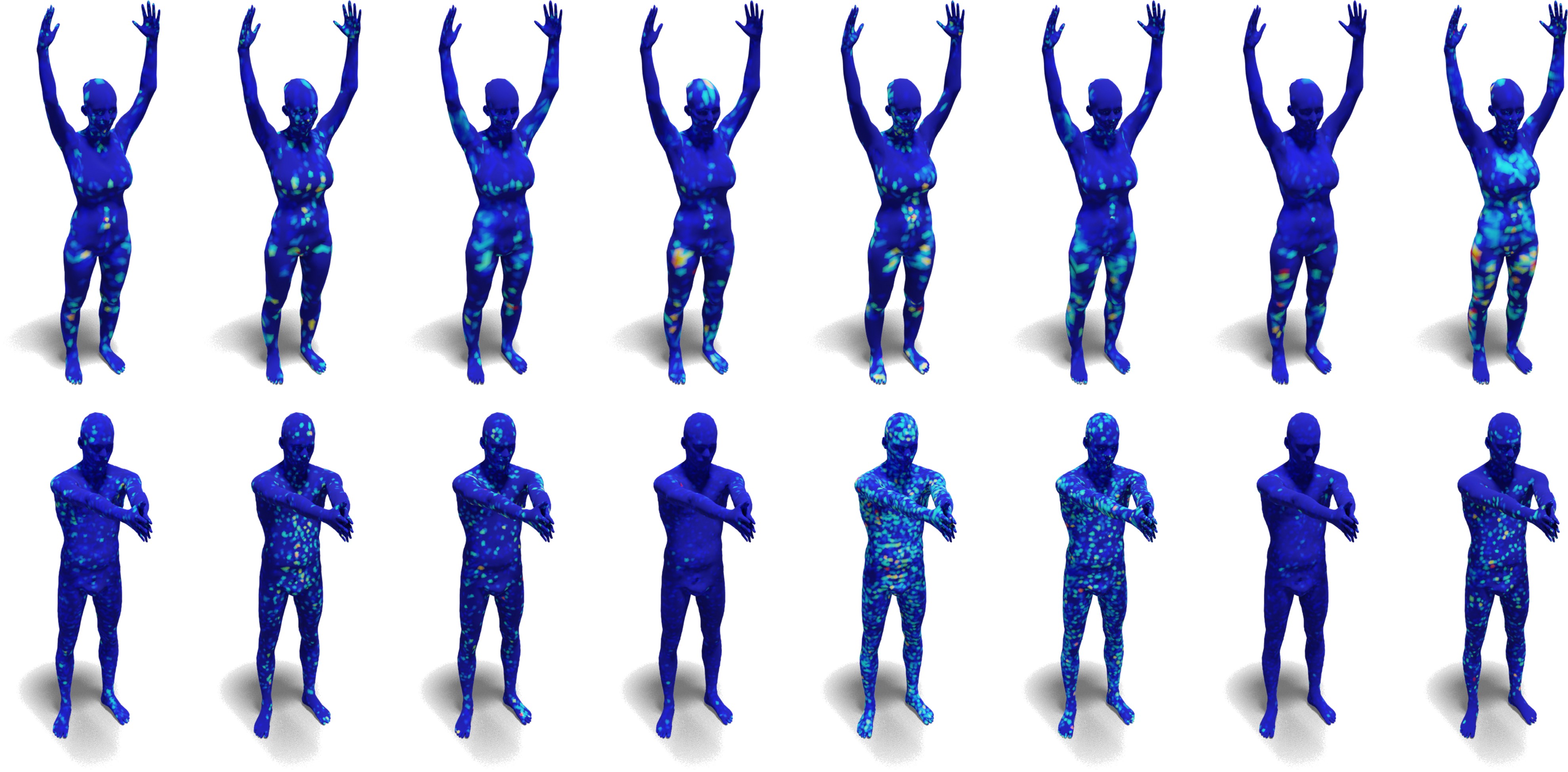}
			\put(40,51){\footnotesize SplineCNN}
			\put(-6,39){\footnotesize Pose 1}
			\put(-6,36){\footnotesize $n$ = 7K}
			\put(-6.5,15){\footnotesize Pose 2}
			\put(-7,12){\footnotesize $n$ = 15K}
			
			\put(3,-1){\footnotesize $d$ = 1}
			\put(18,-1){\footnotesize 37}
			\put(30,-1){\footnotesize 74}
			\put(42,-1){\footnotesize 110}
			\put(55,-1){\footnotesize 147}
			\put(67,-1){\footnotesize 183}
			\put(80,-1){\footnotesize 220}
			\put(93,-1){\footnotesize 256}
		\end{overpic}
		\begin{overpic}[trim=0cm 0.8cm 0cm -2cm,clip,width=0.48\linewidth,grid=false]{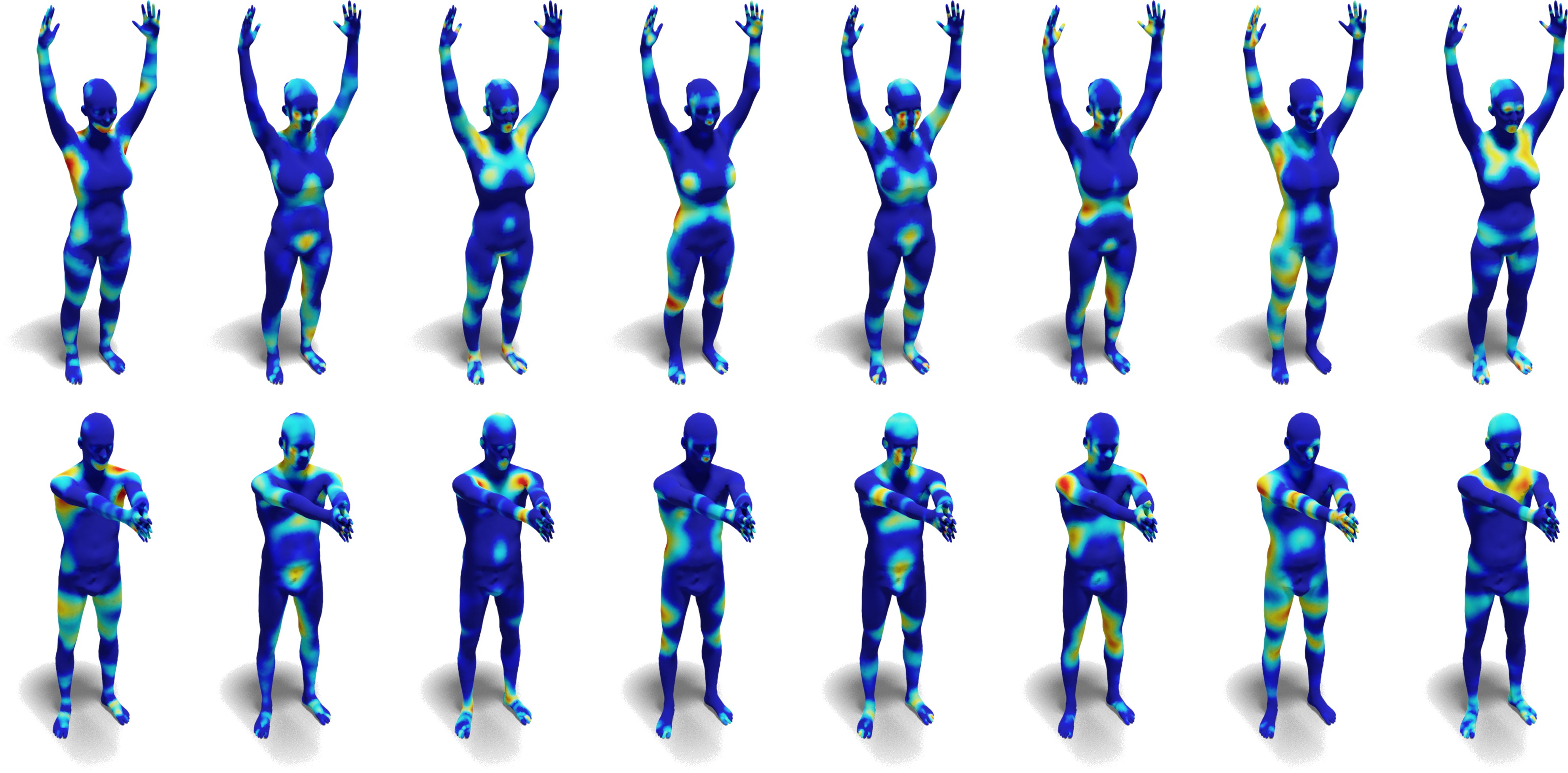}
			\put(40,51){\textbf{\footnotesize MGCN }}
			\linethickness{1pt}
			\multiput(0,8)(0,5){7}{\color{mycolor_dashline}\line(0,1){3}}
		\end{overpic}
		\caption{We train SplineCNN and our MGCN on the WEDS descriptors as visualized in Fig.~\ref{fig:eg:desc_weds} and show the learned descriptors on the same shape and the same dimension. We can see that, though the learned descriptors from SplineCNN has reasonable accuracy in shape matching, the learned descriptors are not smooth and do not encode any semantic information. As a comparison, the learned descriptors from our network are much more coherent between shapes with different poses and resolutions. Also, our learned descriptors are more smooth.}
		\vspace{5pt}
		\label{spline_vs_mgcn}
	\end{figure*}
	
	\begin{figure*}[!t]
		\centering
		\begin{overpic}[trim=0cm 0cm 0cm 0cm,clip,width=1\linewidth,grid=false]{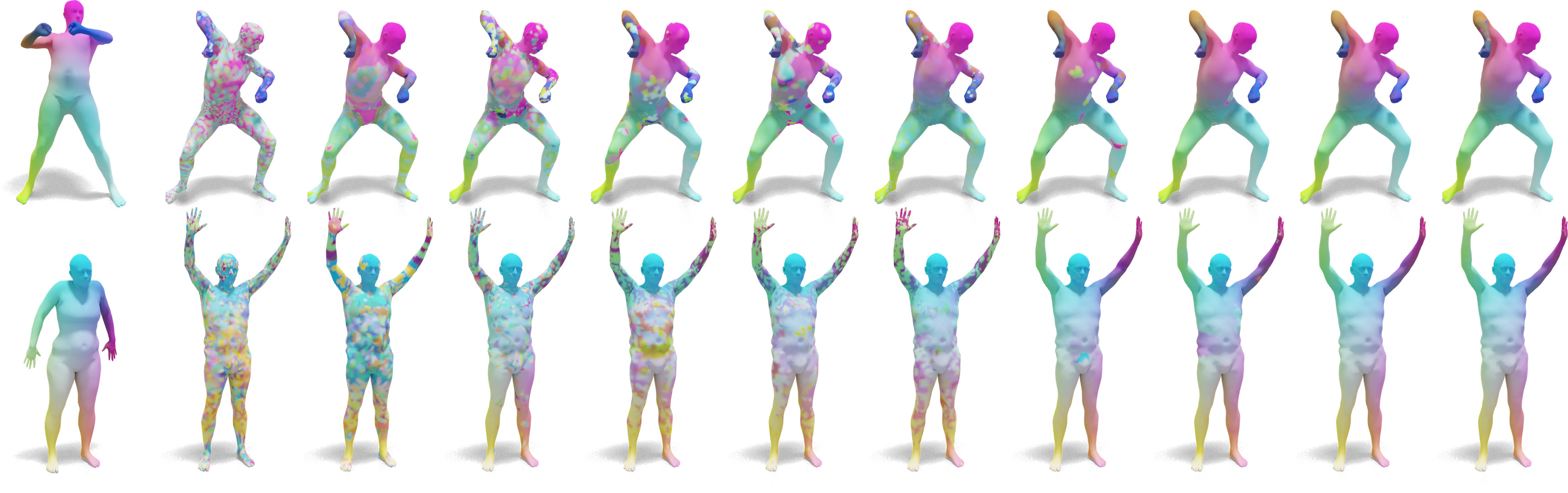}
			\put(3,32){\footnotesize Source}
			\put(12,32){\footnotesize CGF}
			\put(21,32){\footnotesize OSD}
			\put(30,32){\footnotesize SplineCNN}
			\put(39,32){\footnotesize Geo(WKS)}
			\put(48,32){\footnotesize Geo(LPS)}
			\put(57,32){\footnotesize Geo(WEDS)}
			\put(66,32){\footnotesize MGCN(WKS)}
			\put(75,32){\footnotesize MGCN(LPS)}
			\put(83,32){\footnotesize Cheby(WEDS)}
			\put(92,32){\footnotesize \textbf{MGCN(WEDS)}}
		\end{overpic}\vspace{-10pt}
		\caption{Here we show an example pair from SCAPE (top row) and FAUST (bottom row) , where the source and the target shape has the same resolution and we compare the maps obtained from different methods visualized by color transfer. The descriptor input of a network is shown inside brackets.}
		\vspace{5pt}
		\label{fig:faust_scape_matching}
	\end{figure*}

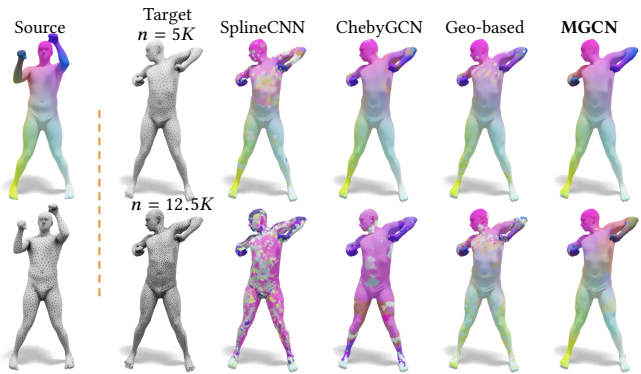
\begin{figure}[!t]
	\centering
	\begin{overpic}[trim=0cm 0cm 0cm 0cm,clip,width=1\linewidth,grid=false]{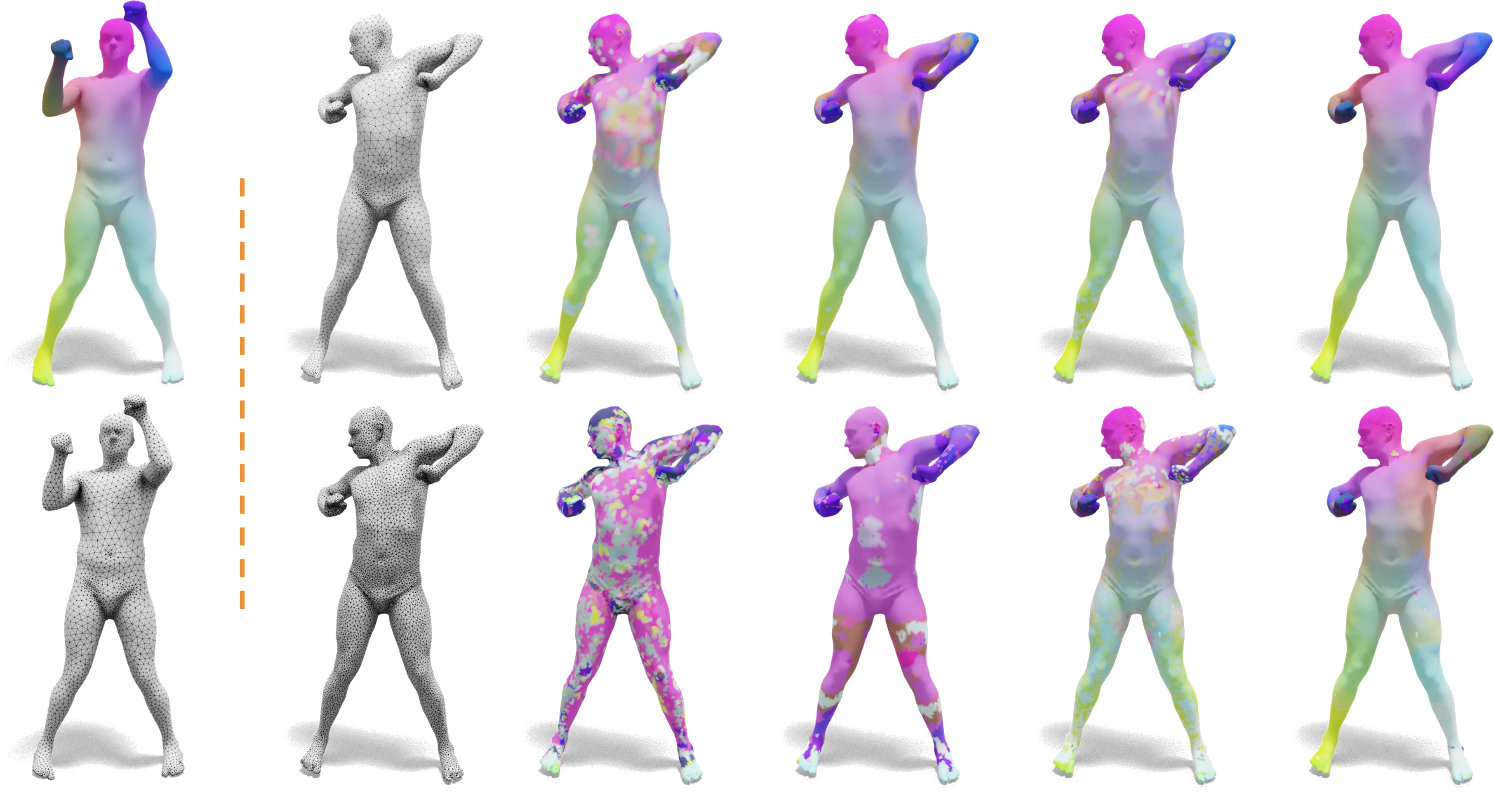}
		\put(3,54.5){\footnotesize Source}
		\put(23,56.5){\footnotesize Target}
		\put(22,53.5){\footnotesize $n=5K$}
		\put(21,27){\footnotesize $n=12.5K$}
		\put(35,54.5){\footnotesize SplineCNN}
		\put(53,54.5){\footnotesize ChebyGCN}
		\put(70,54.5){\footnotesize Geo-based}
		\put(88,54.5){\footnotesize \textbf{MGCN}}
	\end{overpic}\vspace{-3pt}
	\caption{Here we compare the performance of different methods with respect to different resolutions. The source shape has 5K vertices and the target shape has 5K vertices in the top row and 12K vertices in the bottom row. All the networks take WEDS as input. Note that the matches change considerably from the top row to the bottom row for all networks except our MGCN. This indicates that our network can greatly improves upon the generalization performance of the current state of the art.}\vspace{-12pt}
	\label{fig:scape:diff_resolution}
\end{figure}
	
	\subsubsection{Learning descriptor on near-isometric shapes}
	In this task, we focus on learning robust descriptors on different resolutions. \revised{We use two datasets, FAUST and SCAPE, for evaluation that have been remeshed with different algorithms. FAUST has been remeshed by local remeshing operations~\cite{wang2018isotropic} that maintain the positions of the original vertices and SCAPE has been remeshed to about 5K vertices per shape using the LRVD remeshing method~\cite{yan2014low}}.
	The vertex position and triangulation are totally different. We have a ground-truth correspondence between the remeshed vertex and exact 5K points sampled from the original dataset.
	There are not many shape descriptor learning approaches considering different resolutions. One of papers used geodesic-based networks embedded LPS~\cite{Wang_2019_CVPR} to achieve the effect of training at one resolution and testing at another resolution \revised{without} too much decline. \revised{In our study,} three \revised{non-learned descriptors} (WKS, LPS, WEDS) are selected. We set feature dimension to 128 for fair comparison.
	For network comparison, we choose the most competitive methods (CGF32~\cite{khoury2017learning}, OSD~\cite{litman2014learning}, SplineCNN($1$-neiberhood)~\cite{fey2018splinecnn} and geodesic-based networks~\cite{Wang_2019_CVPR}). We also build a network stacked by ChebyGCN ($k$-neiberhood)~\cite{defferrard2016convolutional} layers. The structure is followed by Section~\ref{sec:Architecture}, we replace MGCONV layers with ChebyGCN layers. The rest of the structure such as input and output dimension is the same as MGCN.

	\begin{figure*}[!t]
	\centering
	\begin{overpic}[trim=0cm 0cm 0cm 0cm,clip,width=1\linewidth,grid=false]{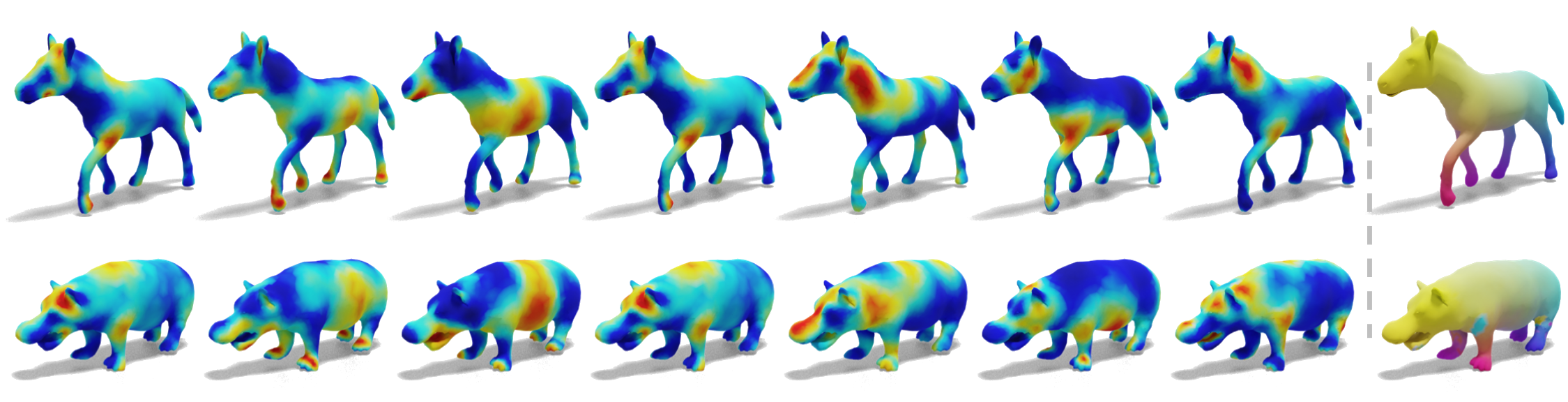}
		\put(95,22){ Source}
		\put(95,10){ Target}
		\put(84,25){ Our map (NN-search)}
		\put(36,25){ Our learned descriptors }
	\end{overpic}\vspace{-12pt}
	\caption{\revised{Here we show an example pair from the SMAL dataset, where the source and the target shape have large deformation. The learned descriptors are shown on the left. We also show the map visualized by color transfer on the right.} }
	\label{fig:horse_hippo}
\end{figure*}
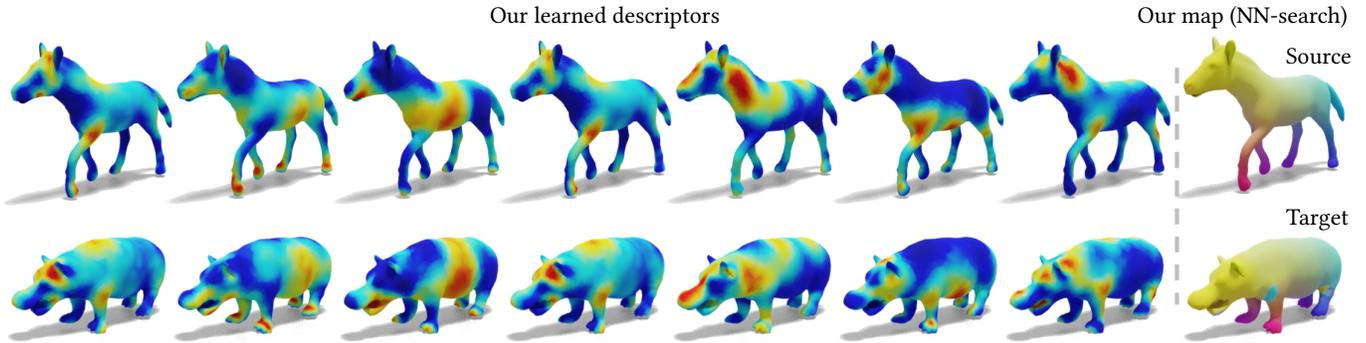

	\paragraph{Experimental results on FAUST}
	In this setting, all the combinations for learning descriptors are generated by learning only on FAUST 6890 vertices and testing on FAUST 6890 and other vertices. We use the nearest neighbor of the feature space to detect the matching discrimination of descriptors between different resolutions. Table~\ref{tab:learn_faust} shows the average geodesic errors on different settings. It can be seen that the ChebyGCN network significantly overfits and the performance drops by a factor of about 100 if the resolution changes.  SplineCNN was designed to directly learn a mapping between points of different shapes, so that we had to make modifications to use SplineCNN for descriptor learning: we take the output of the second to last linear mapping layer as the descriptor to learn. In general, we observe that the performance of SplineCNN, CGF, and OSD is significantly worse than our MGCN.
	Based on the results we consider the geodesic-based network to be our main competitor. Similar to our network, the geodesic-based network is fairly robust to different surface discretizations, however, overall our results are much stronger. For example, the direct error is 3 to 18 times lower when our network is used accross different resolutions.
	Fig.~\ref{fig:faust_learn} reports the curves for the CMC and CGE metric of different descriptors. Therefore, MGCN utilizes GCN's powerful fitting capabilities and guarantees robustness at different resolutions simultaneously. In addition, using different descriptors as input will also affect the network. The results show that the performance of the setting of WEDS and MGCN is the best.
	
	\setlength{\tabcolsep}{1.7em}
	\begin{table}[!t]
		\caption{\revised{Comparison with deep functional maps (FMnet). Here we report the average (direct/symmetry-aware) geodesic error computed on 15$\times$14 shape pairs from FAUST and 10$\times$9 shape pairs from SCAPE.}}\vspace{-3pt}
		\label{tab:mapping}
		\footnotesize
		\begin{tabular}{|l|c|c|}
			\hline
			\multirow{2}{*}{Methods} & \multicolumn{2}{c|}{Dataset} \\ \cline{2-3}
			& FAUST(6890) & SCAPE(12.5K) \\ \hline
			FMNet(desc) + NN & 169 / 58 & 205 / 93 \\
			FMNet & 17 / 11 & 131 / 72 \\
			FMNet + ZoomOut & 15 / 10 & 50 / 36 \\ \hline
			MGCN + NN & \textbf{8 / 7} & \textbf{46 / 16} \\
			MGCN + NN + ZoomOut & \textbf{8 / 6} & \textbf{22 / 14} \\ \hline
		\end{tabular}\vspace{-3pt}
	\end{table}
	
	\paragraph{Experimental results on SCAPE}
	The same network may behave differently on different datasets, we need another dataset to test the network.
	In this setting, all the combinations for learning descriptors are generated by learning only on SCAPE 5K vertices and testing on SCAPE 5K and 12.5K vertices. We then perform feature matching between different resolutions. SCAPE 5K and 12.5K have a completely different shape structure, which is more difficult. Table~\ref{tab:learn_scape} shows average geodesic errors on different settings. Compared with FAUST dataset, OSD has better results on SCAPE, but it seems that overfitting is more severe at resolutions. The descriptors of CGF still perform poorly on the mesh. Same as FAUST, SplineCNN and ChebyGCN overfit at one resolution. Geodesic-based method seems to be more stable with the change of resolution on FAUST, but the discrimination of descriptor is difficult to improve further. The setting of WEDS and MGCN still generates the most discriminative descriptor, while ensuring robustness to the change of resolution. Fig.~\ref{fig:scape_learn} shows the curves for the CMC and CGE metric of different descriptors. Compared with SplineCNN and ChebyGCN, MGCN outputs better performing descriptors that are more robustness at different resolutions. This is also illustrated by Fig.~\ref{spline_vs_mgcn}, where we visualize the learned descriptors of SplineCNN and MGCN.
	
	Fig.~\ref{fig:faust_scape_matching} shows a qualitative example of a pair of FAUST shapes and SCAPE shapes, where we use the learned descriptors to find correspondences and compare the quality of the obtained maps between different competitors.  Another comparison of the performance of different methods with respect to different resolutions is given in Fig.~\ref{fig:scape:diff_resolution}. In both figures, we observe that our learned descriptor generated by MGCN leads to the best maps.

	\paragraph{Comparison with deep functional maps.}
	\revised{Deep functional maps (FMNet)~\cite{litany2017deep} are a state-of-the-art shape matching framework. While the final output of the proposed framework is dense correspondences between a pair of shapes, the first part of the proposed network can be used to learn shape descriptors. Since this competing descriptor learning method is resolution dependent, we only compare on a dataset where all meshes have exactly the same number of vertices. We selected the original FAUST and SCAPE datasets for this comparison (see Table~\ref{tab:mapping}). Specifically, we compare to three settings of FMNet: (1) we apply nearest-neighbor search to the \emph{learned descriptors} from the FMNet to obtain the pointwise correspondence (called "FMNet(desc) + NN" in Table~\ref{tab:mapping}). (2) the direct output of the FMNet (called "FMNet"). (3) we apply ZoomOut~\cite{melzi2019zoomout}, a recent state-of-the-art refinement technique to further refine the output of FMNet.
		Table~\ref{tab:mapping} shows that our method outperforms FMNet in all three different settings. In future work, it will be interesting to compare to an extension of the deep functional maps that was developed concurrently to our work~\cite{donati2020deep}.}

	\setlength{\tabcolsep}{1.1em}
	\begin{table}[!t]
		\caption{\revised{Average direct geodesic error computed on non-isometric animal shapes from SMAL dataset. Our method outperforms both BIM and SplineCNN, the current state-of-the-art non-learning and learning method, respectively.}}\vspace{-3pt}
		\label{tab:non-isometric}
		\footnotesize
		\begin{tabular}{|c|c|c|c|}
			\hline
			Shape Pairs & BIM & SplineCNN + NN & MGCN + NN \\ \hline
			\textbf{Average (8$\times$7 pairs)} & 59 & 225 & \textbf{45} \\ \hline
			Cow, Wolf & 37 & 219 & \textbf{18} \\ \hline
			Tiger, Dog & 39 & 227 & \textbf{38} \\ \hline
			Fox, MaleLion & 104 & 221 & \textbf{51} \\ \hline
			Horse, Hippo & \textbf{76} & 244 & 80 \\ \hline
		\end{tabular}\vspace{-7.5pt}
	\end{table}

	\subsubsection{Learning descriptors on non-isometric shapes}
	
	\revised{We also test our learning framework on a small animal dataset SMAL to see how our pipeline can be generalized to shape pairs with larger deformation. As introduced previously, the SMAL dataset has only ~50 shapes from different animal categories with different poses. Since this number is too small for efficient network training, we use the as-rigid-as-possible~\cite{sorkine2007rigid} deformation method to efficiently extend the dataset to 116 shapes with more poses. We divide the 5K remeshed dataset into 100 shapes for training, 8 shapes for validation, and 8 shapes for testing. The test data set contains shape pairs from 8 different categories to ensure that we evaluate only on non-isometric pairs. We compute the geodesic error for 8 $\times$ 7 non-isometric pairs.
		We consider BIM~\cite{kim2011blended} and SplineCNN~\cite{fey2018splinecnn} as a baseline.  BIM is the current state of the art for non-isometric shape matching and can handle large deformations.

		As before, we use nearest-neighbor matching to evaluate the performance of two learned descriptors. We use WEDS as input for MGCN training. For BIM, we use the complete pipeline as BIM does not compute descriptors.
		Table~\ref{tab:non-isometric} reports the average direct error of 56 non-isometric SMAL shape pairs and we also show the performance of different methods on four example shape pairs with large distortion. It can be seen that MGCN outperforms BIM. Compared with SplineCNN, the performance of our descriptors is significantly better. Fig.~\ref{fig:horse_hippo} shows an example pair of a horse and a hippo. While the map computed by nearest-neighbor search has reasonable quality overall, there are also some outliers. An interesting direction of future work will be to combine our learned descriptors with a suitable optimization method to refine the initial matches. A drawback of our comparison is that we only used a small dataset. In future work, it would be important to collect larger datasets containing correspondences of non-isometric shape pairs.}

	\setlength{\tabcolsep}{1.1em}
	\begin{table}[!t]
		\caption{Average (direct/symmetry-aware) geodesic error computed on 6$\times$5 shape pairs of FAUST. \revised{We vary the number of eigenfunctions (\#basis) and set the parameters \#scales and \#samples to 96}.}\vspace{-3pt}
		\label{tab:eigen1}
		\footnotesize
		\begin{tabular}{|c|c|c|c|c|}
			\hline
			Descriptors & \multicolumn{4}{c|}{\#Basis} \\ \cline{2-5}
			(\#96) & 50 & 100 & 200 & 300 \\ \hline
			HKS & 500 / 382 & \textbf{500 / 381} & 501 / 382 & 501 / 382 \\ \hline
			WKS & 341 / 152 & \textbf{325 / 112} & 358 / 139 & 339 / 118 \\ \hline
			DTEP & 374 / 208 & 302 / \textbf{95} & \textbf{295} / 100 & 366 / 146 \\ \hline
			WEDS & 352 / 110 & 266 / 83 & 263 / \textbf{68} & \textbf{256} / 77 \\ \hline
		\end{tabular}
	\end{table}
	
	\setlength{\tabcolsep}{1.1em}
	\begin{table}[!t]
		\caption{Average (direct/symmetry-aware) geodesic error computed on 6$\times$5 shape pairs of FAUST.
			\revised{We vary the number of eigenfunctions (\#basis) and set the parameters \#scales and \#samples to 128.}}\vspace{-3pt}
		\label{tab:eigen2}
		\footnotesize
		\begin{tabular}{|c|c|c|c|c|}
			\hline
			Descriptors & \multicolumn{4}{c|}{\#Basis} \\ \cline{2-5}
			(\#128) & 50 & 100 & 200 & 300 \\ \hline
			HKS & 503 / 383 & \textbf{ 502 / 383} & 501 / 382 & 502 / 384 \\ \hline
			WKS & 334 / 134 & \textbf{ 322 / 112} & 336  / 126 & 355 / 115 \\ \hline
			DTEP & 397 / 226 & \textbf{ 307 / 96} & 310 / 100 & 377 / 158 \\ \hline
			WEDS & 364 / 126 & 262 / 85 & 257 / \textbf{ 65 } & \textbf{ 248} / 69 \\ \hline
		\end{tabular}
	\end{table}
	
	\subsection{Parameter Settings}

	Different parameters affect the performance of different descriptors. To compare fairly, we need to choose the best parameters for each descriptor. We focus on the spectral descriptors such as HKS, WKS, DTEP. For other descriptors, because of the variety of parameters, we use the parameters recommended by the authors. We analyze the following three parameters of these spectral descriptors: the number of eigenfunctions (\#basis), the number of feature scales (\#scales), and the number of dimensions we sample (\#samples). The number of feature scales in WEDS is $\rm{Num}$, and $\left\lceil {{\rm{\frac{Num}{32}}}}\right\rceil$ denotes the number of wavelet scales we choose to collect the wavelet energy. For other spectral descriptors, the number of feature scales describes how often the time is sampled in the diffusion process. In WKS, the number of scales also encodes diffusion variance. Finally, all feature scales can be chosen to be used in a descriptor. Alternatively, feature scales can be subsampled uniformly. This subsampling is encoded by the parameter \#samples. Therefore, the number of samples has to be smaller or equal to the number of feature scales.
	We test on 6 models resulting in 6$\times$5 shape pairs to evaluate the parameters in the following experiments.

	\paragraph{The number of eigenfunctions.}
	We evaluate four different parameter settings for the number of eigenfunctions (\#basis): 50, 100, 200, and 300. We do this for two settings of the parameter feature scales: 96 and 128. The number of samples is equal to the number of feature scales.
	Table~\ref{tab:eigen1} and Table~\ref{tab:eigen2} show the average geodesic error for these tests. HKS does equally well for all tests. WKS works best with 100 eigenfunctions and the performance descreases with more eigenfunctions. DTEP works well with 100 and 200 eigenfunctions and the performance decreases with 300. WEDS seems to do well with 200 to 300 eigenfunctions.
	For the subsequent tests we pick HKS:100, WKS:100, DEP: 100, WEDS: 300 as the number of eigenfunctions.
	Generally, it can be seen that WEDS has better performance with more eigenfunctions, while other frequency-domain descriptors do not perform better with more eigenfunctions. One possible explanation is that when constructing the WEDS descriptor, the vertex information needs to be reconstructed by the basis functions, so more basis functions may lead to higher reconstruction accuracy.
	
	\setlength{\tabcolsep}{1.3em} 
	\begin{table}[!t] 
		\caption{Average (direct/symmetry-aware) geodesic error computed on 6$\times$5 shape pairs of FAUST.  \revised{We fix the number of eigenfunctions and vary the parameters \#scales and \#samples jointly}.}
		\label{tab:scale}
		\footnotesize
		\begin{tabular}{|l|c|c|c|c|}
			\hline
			\#Scales & HKS & WKS & DTEP & WEDS* \\ \hline
			16 & 508 / 390 & 469 / 290 & 305 / 105 & 290 / 100 \\ \hline
			32 & 499 / 379 & 539 / 371 & 311 / 93 & 269 / 85 \\ \hline
			64 & \textbf{ 495 / 37} & 350 / 142 & 305 / 91 & 257 / 78 \\ \hline
			96 & 500 / 381 & \textbf{ 325 / 112} & \textbf{ 302} / 95 & 256 / 77 \\ \hline
			128 & 502 / 383 & 332 / 112 & 307 / 96 & 248 / 69 \\ \hline
			192 & 506 / 387 & 335 / 117 & 304 / \textbf{ 89} & 247 / 68 \\ \hline
			256 & 510 / 390 & 337 / 120 & 305 / 89 & 245 / 65 \\ \hline
			512 & - & - & - & 244 / 62 \\ \hline
			768 & - & - & - & 245 / 62 \\ \hline
			1024 & - & - & - & \textbf{ 244 / 59} \\ \hline
		\end{tabular}
	\end{table} 
	
	\paragraph{The number of feature scales.}
	We fix the number of eigenfunctions as described before and vary the number of feature scales. We use all feature scales so that the number of samples (\#samples) is equal to the number of feature scales.
	Table~\ref{tab:scale} shows the average geodesic error for different scales. Based on this test, we choose HKS:64, WKS:96, DEP: 96, WEDS: 1024 for the number of scales. Our descriptor behaves differently from other descriptors: the higher the number of feature scales, the better the performance. We believe this is due to the fact that a larger number of feature scales leads to a better accuracy of the signal reconstruction and in turn to a better performance of our descriptor. This behaviour is consistent with the previous test where we also observed that our descriptor performs better with a higher number of eigenfunctions.

	\paragraph{The number of samples.}
	For a fixed number of eigenfunctions and fixed number of feature scales, we vary the parameter \#samples by uniformly subsampling the number of feature scales (\#scales). we show the average geodesic error in the Table~\ref{tab:number} and pick HKS:16(64), WKS:16(96), DEP: 96(96), WEDS: 128(1024). The first number is the parameter \#samples and the number in brackets describes the parameter \#scales. Even though, WEDS has the best performance with 1024 samples, we believe that 128 samples are a better trade-off between memory consumption and descriptor performance. Selecting 1024 samples would also greatly impact the runtime and neural network complexity.

	\setlength{\tabcolsep}{1.2em}
	\begin{table}[!t]
		\caption{Average (direct/symmetry-aware) geodesic error computed on 6$\times$5 shape pairs of FAUST.
			\revised{We fix the number of eigenfunctions and the number of feature scales and vary the parameter \#samples.}}
		\vspace{-6pt}
		\label{tab:number}
		\footnotesize
		\begin{tabular}{|c|c|c|c|c|}
			\hline
			\multirow{2}{*}{Descriptors} & \multicolumn{4}{c|}{\#Samples} \\ \cline{2-5}
			& 16 & 32 & 64 & 96 \\ \hline
			HKS & \textbf{486 / 367} & 489 / 369 & 495 / 375 & - \\ \hline
			WKS & \textbf{323} / 111 & 325 / \textbf{109} & 323 / 111 & 325 / 112 \\ \hline
			DTEP & 309 / 98 & 305 / 96 & 304 / 95 & \textbf{302 / 95} \\ \hline
			\multirow{3}{*}{WEDS} & 295 / 96 & 305 / 103 & 268 / 69 & 286 / 71 \\ \cline{2-5}
			& 128 & 256 & 512 & 1024 \\ \cline{2-5}
			& 250 / 66 & 253 / 59 & 250 / 62 & \textbf{244 / 59} \\ \hline
		\end{tabular}
	\end{table}
	
	\paragraph{Runtime.}
	We also compared the computation time of the four spectral descriptors. The increase in performance is often accompanied by a decrease in computing efficiency. We choose the recommended parameters of these four spectral methods. And the computation cost is shown in Table~\ref{tab:time_cost}. It can be seen that LPS and DTEP have improved performance compared to traditional spectral descriptors, but it takes a lot of time because of the computation of geodesic disks and optimization. WEDS can reduce time consumption while achieving the best performance by energy decomposition.

	\subsection{Limitations and Future Work}
	There are still important challenges left for future work. First, we suspect that our descriptor learning solution still overfits the training data too much, since there is not enough variability in current shape matching datasets. The most beneficial and practical future work would therefore be to collect larger training datasets with more variability.
	Second, \revised{we only use isotropic convolutional kernels in our networks. The anisotropic kernels would be an interesting direction for future work.
		Finally, }our current implementation and evaluation is limited to meshes. It would be interesting to extend our work to point clouds and triangle soups in future work.

	\section{Conclusions}       
	
	We proposed a novel framework for computing two types of shape descriptors. First, the \revised{non-learned} descriptor WEDS is computed using graph wavelets to decompose the Dirichlet energy on a surface. Second, WEDS can be refined by our proposed MGCN to yield a learned descriptor.
	Our results show that the new descriptor WEDS is more
	discriminative than the current state-of-the-art non-learned descriptors and that the combination of WEDS and MGCN is better than the state-of-the-art learned descriptors.
	An important attribute of descriptors is the robustness to different surface discretizations. Our results demonstrate that MGCN generalizes significantly better to different surface discretizations than previous work.

	In this paper, we proposed a descriptor learning framework including
	a new descriptor and a graph neural network. We first verified that WEDS is robust to resolution, rigid transformations, and is also a discriminative descriptor. Then MGCN was proposed to improve the discrimination of \revised{non-learned} descriptors. Most importantly, MGCN can maintain robustness to the change of resolution while improving discrimination. Our framework was demonstrated by comparing to several recent state-of-the-art descriptors and neural networks on near-isometric shapes and non-isometric shapes. Our framework not only improved the performance of the descriptor, but also maintained the robustness to resolution while improving the discrimination of the descriptor.
	
	\begin{acks}
		We would like to thank the anonymous reviewers for their comments.
		This work was supported by the National Key R\&D Program of China (2018YFB2100602 and 2019YFB2204104), the National Natural Science Foundation of China (61620106003, 61772523, 61802406 and 61972459), the Beijing Natural Science Foundation (L182059), the CCF-Tencent Open Research Fund, Shenzhen Basic Research Program (JCYJ20180507182222355), the Alibaba Group through Alibaba Innovative Research Program, and the KAUST OSR Award No. CRG-2017-3426.
	\end{acks}

	\setlength{\tabcolsep}{1.1em}
	\begin{table}[!t]
		\caption{The computation time of different descriptors \revised{with respect to different resolutions}.}\vspace{-6pt}
		\label{tab:time_cost}
		\footnotesize
		\begin{tabular}{|c|c|c|c|c|c|c|}
			\hline
			\multirow{2}{*}{Time(s)} & \multicolumn{6}{c|}{\#Resolution} \\ \cline{2-7}
			& 5K & 6890 & 8K & 10K & 12K & 15K \\ \hline
			HKS & 0.21 & 0.23 & 0.27 & 031 & 0.36 & 0.41 \\ \hline
			WKS & 0.27 & 0.31 & 0.33 & 0.36 & 0.41 & 0.47 \\ \hline
			LPS & 185 & 464 & 655 & 1090 & \multicolumn{1}{c|}{1482} & \multicolumn{1}{c|}{2174} \\ \hline
			DTEP & 1124 & 1467 & 1659 & 2025 & 2235 & 2639 \\ \hline
			WEDS & 16.6 & 24.7 & 42.2 & 65.7 & 82.5 & 144.3 \\ \hline
		\end{tabular}
	\end{table}

	\bibliographystyle{ACM-Reference-Format}
	\bibliography{bibliography}
	
	\vspace{-5pt}
	\appendix
	\section{Appendix}
	
	In this appendix, we give a proof for the reconstruction capabilities of our wavelet filter basis.
	
	\vspace{-5pt}
	\begin{footnotesize}
		\begin{align}
		&\sum\limits_{m{\rm{ = 0}}}^{K} {\sum\limits_v {a\left( v \right)^{-1}{W_{\pmb{\rm{f}}}}\left( {t_m,v} \right)\bm{{\psi} _{t_m,v}}} }  \\
		&= \sum\limits_{m{\rm{ = 0}}}^K {\sum\limits_v {a{{\left( v \right)}^{ - 1}}\sum\limits_{j{\rm{ = 0}}}^{N - 1} {a\left( v \right){g_{{t_m}}}\left( {{\lambda _j}} \right){\sigma _j}{\phi _j}\left( v \right)} \sum\limits_{j{\rm{ = 0}}}^{N - 1} {a\left( v \right){g_{{t_m}}}\left( {{\lambda _j}} \right){\phi _j}\left( v \right){\bm{\phi} _j}} } } \\
		&=\sum\limits_{m{\rm{ = 0}}}^K {\sum\limits_{i{\rm{ = 0}}}^{N - 1} {{g_{{t_m}}}\left( {{\lambda _i}} \right){\sigma _i}} \sum\limits_{j{\rm{ = 0}}}^{N - 1} {{g_{{t_m}}}\left( {{\lambda _j}} \right){\bm{\phi} _j}\sum\limits_v {{\phi _i}\left( v \right)a\left( v \right){\phi _j}\left( v \right)} } } \\
		&=\sum\limits_{m{\rm{ = 0}}}^K {\sum\limits_{j{\rm{ = 0}}}^{N - 1} {g_{{t_m}}^2\left( {{\lambda _j}} \right){\sigma _j}{\bm{\phi} _j}} } \quad (i=j) \\
		&=\sum\limits_{j{\rm{ = 0}}}^{N - 1} {{\sigma _j}{\bm{\phi} _j}} \quad (\mathbf{If} \quad {\sum\limits_{m{\rm{ = 0}}}^{K} {g_{{t_m}}^2\left( {{\lambda _j}} \right)} }=1 \quad \mathbf{Parseval \quad frame}) \\
		&= \pmb{\rm{f}}
		\end{align}
	\end{footnotesize}
\end{document}